\documentclass[aps,prd,superscriptaddress,nofootinbib,tighten,preprint]{revtex4}

\newcommand{\be}{\begin{eqnarray}}
\newcommand{\ee}{\end{eqnarray}}

\def\nue{{\nu_e}}
\def\anue{{\bar\nu_e}}
\def\numu{{\nu_{\mu}}}
\def\anumu{{\bar\nu_{\mu}}}

\newcommand{\ms}{\Delta m^2_{21}}
\newcommand{\ma}{\Delta m^2_{31}}

\def\ltap{\ \raisebox{-.4ex}{\rlap{$\sim$}} \raisebox{.4ex}{$<$}\ }
\def\gtap{\ \raisebox{-.4ex}{\rlap{$\sim$}} \raisebox{.4ex}{$>$}\ }

\def\gs{\mathrel{
   \rlap{\raise 0.511ex \hbox{$>$}}{\lower 0.511ex \hbox{$\sim$}}}}
\def\ls{\mathrel{
   \rlap{\raise 0.511ex \hbox{$<$}}{\lower 0.511ex \hbox{$\sim$}}}}
\newcommand{\bea}{\begin{equation} \begin{array}{c}}
\newcommand{\bead}{\begin{equation} \begin{array}{cccc}}
\newcommand{\eea}{ \end{array} \end{equation}}
\usepackage{amsfonts}
\usepackage{amssymb}
\usepackage{amsmath}
\usepackage{graphicx}
\usepackage{hyperref}
\usepackage{placeins}
\usepackage{gensymb}
\usepackage[T1]{fontenc}
\usepackage{color}
\def\slc#1{\setbox0=\hbox{$#1$}           
    \dimen0=\wd0                                 
    \setbox1=\hbox{/} \dimen1=\wd1               
    \ifdim\dimen0>\dimen1                        
       \rlap{\hbox to \dimen0{\hfil/\hfil}}      
       #1                                        
    \else                                        
       \rlap{\hbox to \dimen1{\hfil$#1$\hfil}}   
       /                                         
    \fi}

\def\nue{{\nu_e}}
\def\anue{{\bar{\nu}_e}}
\def\numu{{\nu_{\mu}}}
\def\anumu{{\bar{\nu}_{\mu}}}

\newcommand{\beq}{\begin{equation}}
\newcommand{\eeq}{\end{equation}}
\newcommand{\beqa}{\begin{eqnarray}}
\newcommand{\eeqa}{\end{eqnarray}}

\newcommand{\tx}{{\theta_{12}}}
\newcommand{\ty}{{\theta_{13}}}
\newcommand{\tz}{{\theta_{23}}}

\newcommand{\da}{\delta_{13}}
\newcommand{\db}{\delta_{24}}

\newcommand{\tmet}{\theta^{3\nu}_{\mu e}}
\newcommand{\tmef}{\theta^{4\nu}_{\mu e}}

\begin{document}

\title{Measuring the Sterile Neutrino CP Phase at DUNE and T2HK}

\author{Sandhya Choubey}
\email{sandhya@hri.res.in}
\affiliation{Harish-Chandra Research Institute, HBNI, Chhatnag Road, Jhunsi, Allahabad 211 019, India}
\affiliation{Department of Physics, School of
Engineering Sciences, KTH Royal Institute of Technology, AlbaNova
University Center, 106 91 Stockholm, Sweden}

\author{Debajyoti Dutta}
\email{debajyotidutta@hri.res.in}
\affiliation{Harish-Chandra Research Institute, HBNI, Chhatnag Road, Jhunsi, Allahabad 211 019, India}

\author{Dipyaman Pramanik}
\email{dipyamanpramanik@hri.res.in}
\affiliation{Harish-Chandra Research Institute, HBNI, Chhatnag Road, Jhunsi, Allahabad 211 019, India}

\begin{abstract}
The CP phases associated with the sterile neutrino cannot be measured in the dedicated short-baseline experiments being built to test the sterile neutrino hypothesis. On the other hand, these phases can be measured in long-baseline experiments, even though the main goal of these experiments is not to test or measure sterile neutrino parameters. In particular, the sterile neutrino phase $\delta_{24}$ affects the charged-current electron appearance data in long-baseline experiment. In this paper we show how well the sterile neutrino phase $\delta_{24}$ can be measured by the next-generation long-baseline experiments DUNE, T2HK (and T2HKK). We also show the expected precision with which this sterile phase can be measured by combining the DUNE data with data from T2HK or T2HKK. The T2HK experiment is seen to be able to measure the sterile phase $\delta_{24}$ to a reasonable precision. We also present the sensitivity of these experiments to the sterile mixing angles, both by themselves, as well as when DUNE is combined with T2HK or T2HKK. 
\end{abstract}
\maketitle

\section{Introduction}

Neutrino oscillation physics has reached precision era. Observation of neutrino oscillations by the solar \cite{PhysRevLett.87.071301} and atmospheric \cite{PhysRevLett.81.1562} neutrinos have been confirmed independently at accelerator \cite{Ahn:2006zza, Michael:2006rx} and reactor \cite{Eguchi:2002dm} experiments. The two mass squared differences and all three mixing angles of the three-generation neutrino oscillation theory have been well determined and the three-generation paradigm well established \cite{Kodama:2000mp}. The only remaining questions which are still to be answered are CP violation in the leptonic sector, neutrino mass hierarchy (that is, whether $\nu_{3}$ is the lightest or the heaviest) and whether $\theta_{23}$ lies in the lower or in the higher octant. Future experiments like DUNE \citep{Acciarri:2016crz,Acciarri:2015uup,Strait:2016mof,Acciarri:2016ooe} and  T2HK \citep{Abe:2015zbg,Abe:2016ero} are going to explore these questions and are expected to come with definite answers.  

Even though neutrino oscillation with three-generations is well established, there are some hints of neutrino oscillations at a higher frequency corresponding to a mass-squared difference $\Delta m^{2} \sim 1$ eV$^{2}$  \citep{Abazajian:2012ys}. LSND experiment \citep{Athanassopoulos:1995iw,Aguilar:2001ty} in Los Alamos, USA, first showed evidence for such oscillations, where a $\anumu$ beam was sent to a detector and the observations showed a 3.8$\sigma$ excess in the positrons, which could be explained in terms of $\anumu\rightarrow\anue$ oscillations. For the $L$ and  $E$ applicable for the LSND experiment, this oscillation corresponds to a mass-squared difference of $\Delta m^{2}\sim 1$ eV$^{2}$.  Experiments like KARMEN \citep{Gemmeke:1990ix} and MiniBooNE \citep{AguilarArevalo:2007it,Aguilar-Arevalo:2013pmq,AguilarArevalo:2010wv} tested the claim. While KARMEN data did not show any evidence for $\anumu\rightarrow\anue$ oscillations, it could not rule out the entire allowed region from LSND. More recently, the MiniBooNE experiment ran in both neutrino as well as antineutrino mode. MiniBooNE did not find  any significant excess in their neutrino mode, however they reported some excess in the antineutrino mode consistent with the LSND result. Apart from these, MiniBooNE also reported some excess in the low energy bins for both neutrino and antineutrino appearance channels, but these cannot be explained in terms of neutrino flavour oscillations. The one additional sterile neutrino can be fitted along with the three active neutrinos in the so-called 3+1  \citep{Goswami:1995yq} type neutrino mass spectrum. The global fit of all the relevant short-baseline data shows severe tension between the appearance and disappearance 
data sets and the overall goodness of fit is only 31 \%.  However, if one consider appearance and disappearance channels separately, the goodness of fit improves slightly and it becomes 50 \% and 35 \%, respectively \citep{Gariazzo:2017fdh}.

In the 3+1 scenario we have 3 mass squared differences\footnote{We define $\Delta m_{ij}^2 = m_i^2 - m_j^2$.} $\Delta m_{41}^2$, $\Delta m_{31}^2$ and $\Delta m_{21}^2$, 6 mixing angles and 3 phases. ln addition to the mixing angles $\theta_{12}$, $\theta_{23}$ and $\theta_{13}$ which appear also  in the three-generation sector we have 3 additional angles $\theta_{14}$, $\theta_{24}$ and $\theta_{34}$ involving the fourth generation. Also, there are two new CP phases $\delta_{24}$ and $\delta_{34}$ in addition to the standard CP phase $\delta_{13}$. At short-baseline experiments, the oscillation probabilities $P_{ee}$, $P_{\mu\mu}$ and $P_{\mu e}$ in the 3+1 scenario can be written in an effective two-generation framework which depends only on $\Delta m_{41}^2$ and an effective mixing term given as a combination of the sterile mixing angles $\theta_{14}$, $\theta_{24}$ and $\theta_{34}$. As a result, the short-baseline experiments are completely insensitive to the sterile CP phases $\delta_{24}$ and $\delta_{34}$. On the other hand, it now well known that even though the $\Delta m_{41}^2$-driven oscillations get averaged out, these phases show up in the oscillation probabilities at the long-baseline experiments. A lot of effort in the last couple of years has gone into estimating the impact of the sterile neutrino mixing angles and phases on the measurement of standard oscillation parameters at long-baseline experiments \citep{Gandhi:2015xza,Dutta:2016glq,Berryman:2015nua,deGouvea:2014aoa,Agarwalla:2016xlg,Agarwalla:2016xxa,Agarwalla:2016mrc,Klop:2014ima, Choubey:2017cba,Ghosh:2017atj}.\footnote{Some recent studies on other new physics scenarios in the context of DUNE and T2HK can be found here \citep{Masud:2015xva,deGouvea:2015ndi,Coloma:2015kiu,Liao:2016hsa,Masud:2016bvp,Agarwalla:2016fkh,Blennow:2016etl,Deepthi:2017gxg,Liao:2016orc,Fukasawa:2016lew,Dutta:2016vcc,Dutta:2016czj,Blennow:2016jkn,Dutta:2016eks,Escrihuela:2016ube,Ge:2016xya,Escrihuela:2015wra,Ge:2016dlx,Tang:2017khg,Choubey:2017dyu,Chatterjee:2015gta,Berryman:2016szd}.}  These papers showed that in presence of sterile neutrino mixing the sensitivity to the measurement of CP violation, mass hierarchy, as well as octant of $\theta_{23}$ becomes a band, where the width of the band comes from the uncertainty on both the values of the sterile mixing angles as well as the sterile phases. While the sterile neutrino mixing angles are constrained by the global short-baseline data, there are no constraints on the sterile phases. In the future, bounds on the sterile neutrino mixing angles are expected to improve by the data from forthcoming short-baseline experiments \cite{Antonello:2015lea,Antonello:2012hf,Jones:2011ci}. Studies have shown that the long-baseline experiments could also give constraints on the sterile neutrino mixing at their near \citep{Choubey:2016fpi} and far \citep{Berryman:2015nua,Kelly:2017kch} detectors. The sterile phases on the other hand, can be constrained {\it only} in the long-baseline experiments. A short discussion on the study of sterile phases were done at T2K+reactor \cite{Klop:2014ima} and T2K+NOvA \citep{Agarwalla:2016mrc} and the sensitivity was shown to be poor. In this paper, we study how well the next-generation experiments DUNE and T2HK will be able to measure the sterile phases. We give the expected sensitivity of DUNE and T2HK alone as well as by combining data from the two experiments. To the best of our knowledge, this is the first time such a complete study is being performed. While the authors in \citep{Agarwalla:2016xxa} did attempt to present the expected precision on the sterile phase (which in their parametrisation was $\delta_{14}$) in the DUNE experiment, the analysis they performed has several short-comings. In their analysis, the authors of \citep{Agarwalla:2016xxa} keep the sterile mixing angles $\theta_{14}$, $\theta_{24}$ and $\theta_{34}$ fixed in the fit at their assumed true values. We allow these angles to vary freely in the fit we perform in this paper. This allows the uncertainty due to both the mixing angles as well as the phases to impact out final results. We also keep our sterile neutrino mixing angles within the currently allowed limits, which have been updated following the results from the NEOS, MINOS and MINOS+ experiments  \citep{Gariazzo:2017fdh}. Expected precision on the sterile phase from T2HK has never been studied before and we present them for the first time. 
We will also show the combined expected sensitivity of DUNE and T2HK to constrain the sterile mixing angles $\theta_{24}$ and $\theta_{14}$, both when the 3+1 scenario is true as well as when there is no positive evidence for sterile neutrino oscillations.

The paper is organised as follows: In section \ref{sec:sim} we briefly describe the sterile neutrino hypothesis and the simulation procedure. In the same section we provide the details of the DUNE, T2HK and T2HKK experiments. In section \ref{sec:phases} we give our main results on how well the future long-baseline experiments will constrain the sterile phase $\db$ if the 3+1 scenario is indeed true. In section \ref{sec:mixangle} we present our results on the expected constraints from future long-baseline experiments if the experiments do not see any positive signal for sterile neutrino oscillations. Finally, we conclude in section \ref{sec:conclusion}.

\section{Sterile Neutrino mixing and simulation \label{sec:sim}}

As discussed above, extra light neutrino state(s) with mass $\sim\mathcal{O}(1$eV$^{2}$) have been proposed to explain the LSND results. In this article we have considered one additional sterile neutrino within the so-called 3+1 scenario, where the three active neutrinos are separated from the sterile neutrinos by a mass gap of $\sim$ 1 eV. In this scenario the mixing matrix is $4\times 4$ and hence is defined in terms of six mixing angles and three phases. The mixing matrix can be parametrised in the following way:
 \begin{equation}
 U^{3+1}_{PMNS} = O(\theta_{34},\delta_{34})O(\theta_{24},\delta_{24})R(\theta_{14})R(\theta_{23})O(\theta_{13},\delta_{13})R(\theta_{12})
 \,.
 \label{eq:umix}
 \end{equation}
Here $O(\theta_{ij},\delta_{ij})$ are 4$\times$4 orthogonal matrices with associated phase $\delta_{ij}$ in the $ij$ sector, and $R(\theta_{ij})$ are the rotation matrices in the $ij$ sector. There are three mass-squared differences in the 3+1 scenario - the solar mass-squared difference $\Delta m_{21}^2 \simeq 7.5 \times 10^{-5}$ eV$^2$, the atmospheric mass-squared difference $\Delta m_{31}^2 \simeq 2.5 \times 10^{-3}$ eV$^2$, and the LSND mass-squared difference $\Delta m_{41}^2 \simeq  1$ eV$^2$. One can of course write the most general oscillation probabilities in terms of these three mass-squared differences, six mixing angles and three phases. However, since oscillation driven by a given mass-squared difference depends on the $L/E$ of the experiment concerned, the expression for the oscillation probabilities simplify accordingly. In particular, the short-baseline experiments have a very small $L/E$ such that $\sin^2(\Delta m_{ij}^2 L/4E) \sim 0$ for $\Delta m_{21}^2$ and $\Delta m_{31}^2$ and only the terms for $\Delta m_{41}^2$ survive. This is the one-mass-scale-dominance case, where only one oscillation frequency due to one mass scale survives. As a result the short-baseline experiments depend on only ``effective'' sterile mixing angles, which are combinations of the mixing angles $\theta_{ij}$ in Eq.~(\ref{eq:umix}). More importantly, since they have only one oscillation frequency, they do not depend on any CP violation phase. Hence, short-baseline experiments are completely insensitive to the sterile phases for the 3+1 scenario.\footnote{In the 3+2 mass spectrum case, there are two sterile neutrinos and two mass squared difference that affect the oscillations at very short baselines. In this case therefore, the short baseline experiments are sensitive to the sterile CP phases.}

In long-baseline experiments such as T2HK and DUNE, the oscillations driven by $\Delta m^{2}_{31}$ dominate while those driven by $\Delta m^{2}_{21}$ are sub-dominant, while the very fast oscillations driven by $\Delta m^{2}_{41}\sim\mathcal{O}(1$eV$^{2}$) get averaged out. 
The transition probability $P_{\mu e}$ in the limit $\sin^2(\Delta m_{41}^2 L/4E) \sim 1/2$ and neglecting earth matter effect is \cite{Gandhi:2015xza}:
\begin{equation}
P_{\mu e}^{4\nu} = P_{1} + P_{2}(\da)+P_{3}(\db)+P_{4}(\da+\db).
\label{eq:pmue}
\end{equation}
Here $P_{1}$ is the term independent of any phase, $P_{2}(\da)$ depends only on $\da$, $P_{3}(\db)$ depends only on $\db$ and $P_{4}(\da+\db)$ depends on the combination ($\da+\db$). The full expression of the different terms in Eq.~(\ref{eq:pmue}) are as follows:
\beqa
P_1 &=& \frac{1}{2}\sin^22\tmef
\nonumber \\
&+& (a^2\sin^22\tmet - \frac{1}{4}\sin^22\ty\sin^22\tmef)
(\cos^{2}\theta_{12}\sin^{2}\Delta_{31}+\sin^{2}\theta_{12}\sin^{2}\Delta_{32})
\nonumber \\ &+& (b^2a^2-\frac{1}{4}a^2\sin^22\tx\sin^22\tmet
- \frac{1}{4}
\cos^4\ty\sin^22\tx\sin^22\tmef) \sin^2\Delta_{21}\,,
\label{eq:p1}
\eeqa
\beqa
P_{2}(\delta_{13}) &=& ba^2  \sin2\tmet 
\big[\cos(\da)\big( \cos2\tx\sin^2\Delta_{21}+\sin^2\Delta_{31}-\sin^2\Delta_{32}\big)\nonumber \\ &-& \frac{1}{2} \sin(\da)
\big(\sin2\Delta_{21}-\sin2\Delta_{31}+\sin2\Delta_{32} \big)\big]\,,
\label{eq:p2}
\eeqa
\beqa
P_{3}(\delta_{24})  &=&   ba \sin2\tmef 
\big[\cos(\db)\big(\cos2\tx\cos^2\ty\sin^2\Delta_{21}-\sin^2\ty(\sin^2\Delta_{31}
-\sin^2\Delta_{32}\big) \big)
\nonumber \\ &+& \frac{1}{2} \sin(\db) 
\big(\cos^2\ty\sin2\Delta_{21}+\sin^2\ty(\sin2\Delta_{31}-\sin2\Delta_{32})
\big)\big]\,,
\label{eq:p3}
\eeqa
\beqa
P_{4} (\delta_{13} + \delta_{24})  &=&  a \sin2\tmet \sin2\tmef
\big[\cos(\da + \db) \big(-\frac{1}{2}\sin^22\tx\cos^2\ty\sin^2\Delta_{21} 
\nonumber \\ &+& \cos2\ty(\cos^2\tx\sin^2\Delta_{31}+\sin^2\tx\sin^2\Delta_{32})\big)
\nonumber \\ &+& \frac{1}{2} \sin(\da + \db) 
\big(\cos^2\tx\sin2\Delta_{31}+\sin^2\tx\sin2\Delta_{32} \big)\big]\,,
\label{eq:p4}
\eeqa
where,
\beqa 
\sin2\theta_{\mu e}^{3\nu}=\sin2\ty\sin\tz\nonumber  \\
b=\cos\theta_{13}\cos\theta_{23}\sin2\tx \nonumber  \\
\sin2\theta_{\mu e}^{4\nu}=\sin2\theta_{14}\sin\theta_{24} \nonumber \\
a=\cos\theta_{14}\cos\theta_{24}.
\end{eqnarray}
We can see from Eq.~(\ref{eq:pmue}) that even though the $\Delta m^{2}_{41}$-driven oscillations are averaged out, the CP phases associated with the sterile sector still appear in the neutrino oscillation probability $P_{\mu e}$. This dependence comes in term $P_3(\delta_{24})$ that depend only on the sterile phase $\delta_{24}$ as well as in term $P_{4} (\delta_{13} + \delta_{24})$ which depends on combination of $\delta_{13}$ and $\delta_{24}$. Hence, we can expect the next-generation long-baseline experiments to be sensitive to the sterile phases. We will see the anti-correlation between $\delta_{13}$ and $\delta_{24}$ manifest in our results on measurement of these phases in the long-baseline experiments. In fact, as has been pointed above, the sterile CP phases cannot be measured in the short-baseline experiments which are dedicated to measuring the sterile neutrino mixing. Hence, experiments like DUNE and T2HK are the only place where $\delta_{24}$ can be measured in the 3+1 scenario. Note that in Eq.~(\ref{eq:pmue}) the probability $P_{\mu e}$ does not depend on the mixing angle $\theta_{34}$, hence the corresponding phase associated with this angle $\delta_{34}$ also does not appear. Once earth matter effects are taken into account the probability $P_{\mu e}$ picks up a $\theta_{34}$ dependence and hence depends on $\delta_{34}$ as well. However, for DUNE and T2HK experiments earth matter effects are rather weak and hence their corresponding sensitivity to $\delta_{34}$ cannot be expected to be strong. Therefore, as we will see in the Results section, these experiments are mainly able to put constraints on $\delta_{24}$.

 Prior to proceeding, we briefly discuss our simulation procedure as well as the present statues of the neutrino oscillation parameters. For our analysis we have used GLoBES (Global Long Baseline Experiment Simulator) \citep{Huber:2004ka,Huber:2007ji} along with the additional codes \citep{Kopp:2006wp,Kopp:2007ne} for calculating probabilities in the 3+1 scenario. We have used constant matter density for all the cases. Throughout the analysis we choose the true values\footnote{Throughout this paper we refer to the oscillation parameter values at which the ``data" is generated as the ``true value" and values in the fit as ``test values".} of the standard oscillation parameters as: $\theta_{12} = 33.56\degree$, $\theta_{13} = 8.46\degree$, $\theta_{23} = 45\degree$, $\ms = 7.5 \times 10^{-5}$ eV$^{2}$, $\ma = 2.5 \times 10^{-5}$ eV$^{2}$ and $\delta_{13} = -90^\circ$ unless specified otherwise. This choice of parameters are consistent with the current limits \citep{Esteban:2016qun}.  Although one should marginalize over all the free parameters whenever one introduces some new physics, but new physics scenarios often give large number of parameters. Marginalisation over these large number of parameters is computationally challenging, so one has to do some approximation. For our analysis we have checked that the effect of marginalisation over the standard three neutrino parameters other than $\delta_{13}$ have no significant effect.  So to save computational time we did not marginalise over these not so relevant set of parameters.

For the sterile neutrino mixing, we have considered two scenarios. We first start by assuming that active-sterile neutrino oscillations  indeed exist and find the expected constraints on the sterile neutrino mixing angles and phase $\db$ assuming non-zero sterile neutrino mixing angles in the ``data''. For this case we generate the ``data'' at true values $\Delta m^{2}_{41} \approx 1.7$ eV$^{2}$, $\theta_{14} \approx 8.13\degree$, $\theta_{24} \approx 7.14\degree$, $\theta_{34} = 0\degree$, which are the current best-fit values taken from \citep{Gariazzo:2017fdh}. The true values of the sterile phases will be specified. We marginalise our $\chi^2$ over all sterile mixing parameters except $\Delta m^{2}_{41}$ in the fit. The $\chi^2$ is marginalised by varying $\theta_{14}$, $\theta_{24}$ and $\theta_{34}$ in the range [$5\degree$, $10.5\degree$], [$4\degree$, $9.5\degree$], and [$0\degree$, $12\degree$], respectively \citep{Gariazzo:2017fdh} without any Gaussian prior, while the phases $\delta_{24}$ and $\delta_{34}$ are varied in their full range $\in [-180\degree, 180\degree$].

We next assume a scenario where the sterile neutrino mixing does not exist in nature and we show how well would then the long-baseline experiments DUNE and T2HK constrain the sterile neutrino mixing angles. For this case the data of course corresponds to true sterile mixing angles zero. The marginalisation of the $\chi^2$ is done over all the three sterile mixing angles and the three phases. Mixing angles  $\theta_{14}$, $\theta_{24}$ and $\theta_{34}$ are marginalised in the range [$0\degree$, $10\degree$], [$0\degree$, $10\degree$], and [$0\degree$, $12\degree$], respectively, while the phases are allowed to vary in their  full range $\in [-180\degree, 180\degree$].
\subsection{DUNE}

DUNE (Deep Underground Neutrino Experiment) \citep{Acciarri:2016crz,Acciarri:2015uup,Strait:2016mof,Acciarri:2016ooe} is a future long baseline experiment proposed in US. Purpose of DUNE is to address all the three unknowns of the neutrino oscillation sector - the leptonic CP violation, mass-hierarchy and the octant of the $\theta_{23}$. DUNE will consist of a source facility at Fermilab and a far detector at Sanford Underground Research facility in South Dakota at a distance of 1300 km from the source. Hence, the baseline of the experiment is 1300 km. The accelerator facility at Fermilab will give a proton beam of energy 80-120 GeV at 1.2-2.4 MW which will eventually give a wide-band neutrino beam of energy range 0.5-8.0 GeV. The far site will consist of 4 identical detector of 10 kt each which will give fiducial mass of 34 kt. All the detectors will be LArTPC (Liquid Argon Time Projection Chamber).

In this work we have considered a Liquid Argon detector of fiducial mass 34 kt at a baseline of 1300 km. The neutrino flux is given by the 120 GeV, 1.2 MW proton beam. Here we have considered 5 years of neutrino and 5 years of antineutrino. Appearance and disappearance channels are combined for the analysis. The energy resolutions for the $\mu$ and $e$ are taken to be  $20\%/\sqrt{E}$ and $15\%/\sqrt{E}$, respectively. The signal efficiency is taken to be 85\%. The backgrounds are taken from \citep{Acciarri:2015uup}. In the neutrino (antineutrino) mode, the signal normalization error is 2\% (5\%), the background normalization error is 10\% (10\%) and the energy calibration error is 5\% (5\%). This choice of systematics is conservative compared to the projected systematics in \citep{Acciarri:2015uup}. \footnote{The experimental specifications for DUNE has been updated in \cite{Alion:2016uaj}. However, we have explicitly checked that the physics results do not differ much for the two specifications. The older version was for 5+5 years of run while the newer one is for 3.5+3.5 years of run and yet inspite of lower statistics the newer version is able to achieve similar physics goal because of the optimised fluxes, detector response and systematics.}

\subsection{T2HK}

The Hyper-Kamiokande (HK) \citep{Abe:2015zbg,Abe:2016ero} is the upgradation of the Super-kamiokande (SK) \citep{Fukuda:1998mi} program in Japan, where the detector mass is projected to be increased by about twenty times the fiducial mass of SK. HK will consist of two 187 kt water Cherenkov detector modules, to be place near the current SK site about 295 km away from source. The detector will be  2.5$\degree$ off-axis from the J-PARC beam which is currently being used by the T2K experiment \citep{Abe:2011ks}. T2HK has similar physics goals as DUNE, but since it will employ a narrow-band beam, it can be complimentary to the DUNE experiment. 

For our analysis we take a beam power of 1.3 MW and the 2.5$\degree$ off-axis flux. We consider a baseline of 295 km and the total fiducial mass of 374 kt (two tank each of which is 187 kt). We consider 2.5 years of neutrino and 7.5 years of anti-neutrino runs in both appearance and disappearance channels. The energy resolution is taken to be $15\%/\sqrt{E}$. The number of events are matched with the TABLE II  and TABLE III of \citep{Abe:2016ero}. The signal normalization error in $\nue(\anue)$ appearance and $\numu(\anumu)$ disappearance channel are  3.2\% (3.6\%) and 3.9\% (3.6\%), respectively. The background and energy calibration errors in all channels are 10\% and 5\%, respectively.

\subsection{T2HKK}

In \citep{Abe:2016ero}, the collaboration has also discussed the possibility of shifting one of the water tanks to a different location in Korea at a distance of about 1100 km from the source. This proposed configuration will consist of the same neutrino source but with one detector of fiducial mass 187 kt at a baseline of 295 km and another similar detector of fiducial mass  187 kt at a baseline of 1100 km. The second oscillation maximum takes place near $E_{\nu}= 0.6$ GeV at the second detector. Both detectors are taken 2.5$\degree$ off-axis in our study. Since the flux peaks at the same energy for both detector locations, the Japan detector sees the flux at the first oscillation maximum while the Korea detector sees it at the second oscillation maximum. This whole setup is called T2HKK. In our analysis, we have considered signal normalization error of 3.2\% (3.6\%) in $\nue$ $(\anue)$ appearance channel and 3.9\% (3.6\%) in  $\numu$ $(\anumu)$ disappearance channel, respectively. The background and energy calibration errors  are 10\% and 5\% in all the channels, respectively.

\section{Measurement of the sterile phases\label{sec:phases}}

\begin{figure}{}
\includegraphics[width=0.45\textwidth]{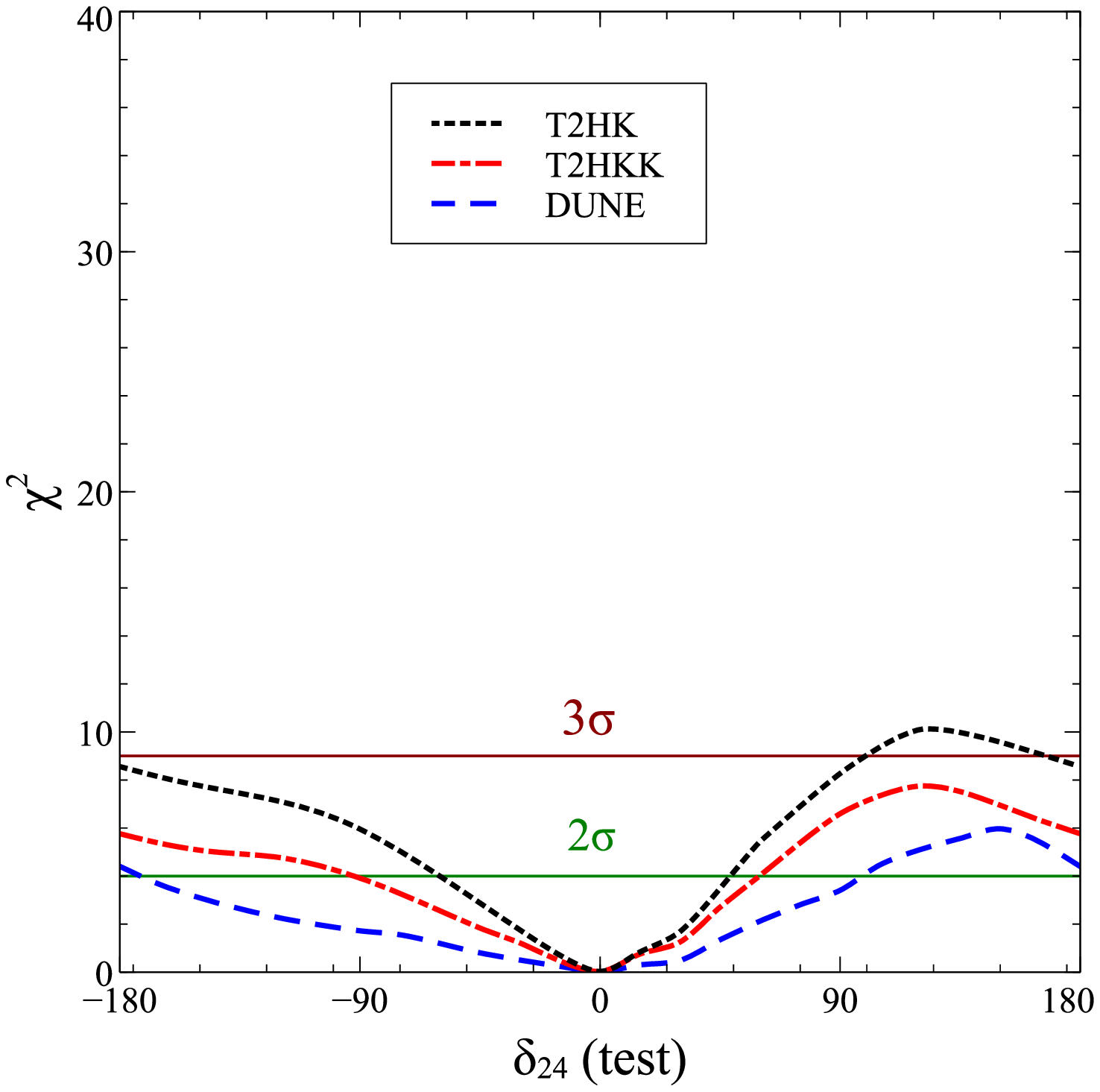}
\includegraphics[width=0.45\textwidth]{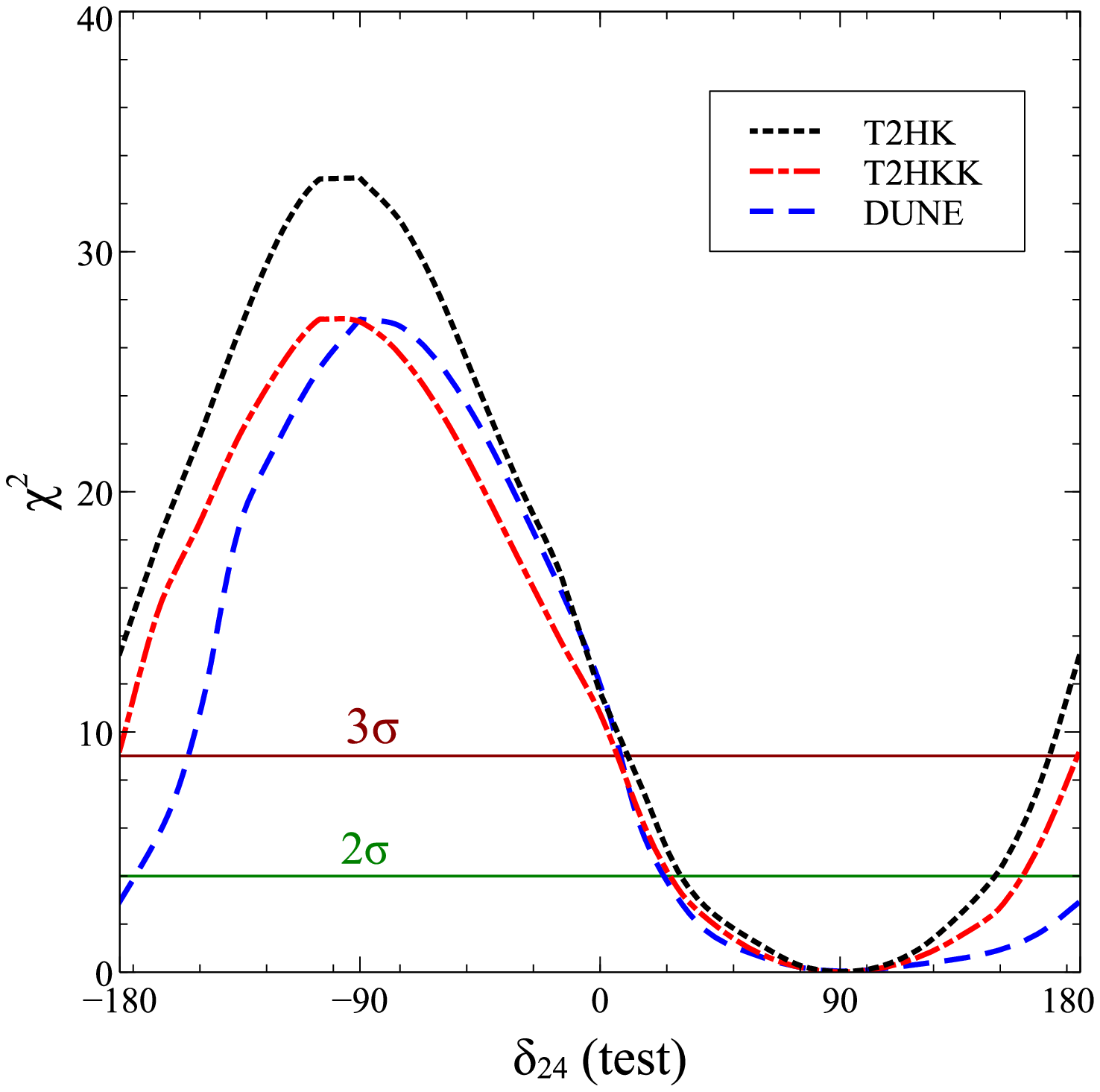}
\includegraphics[width=0.45\textwidth]{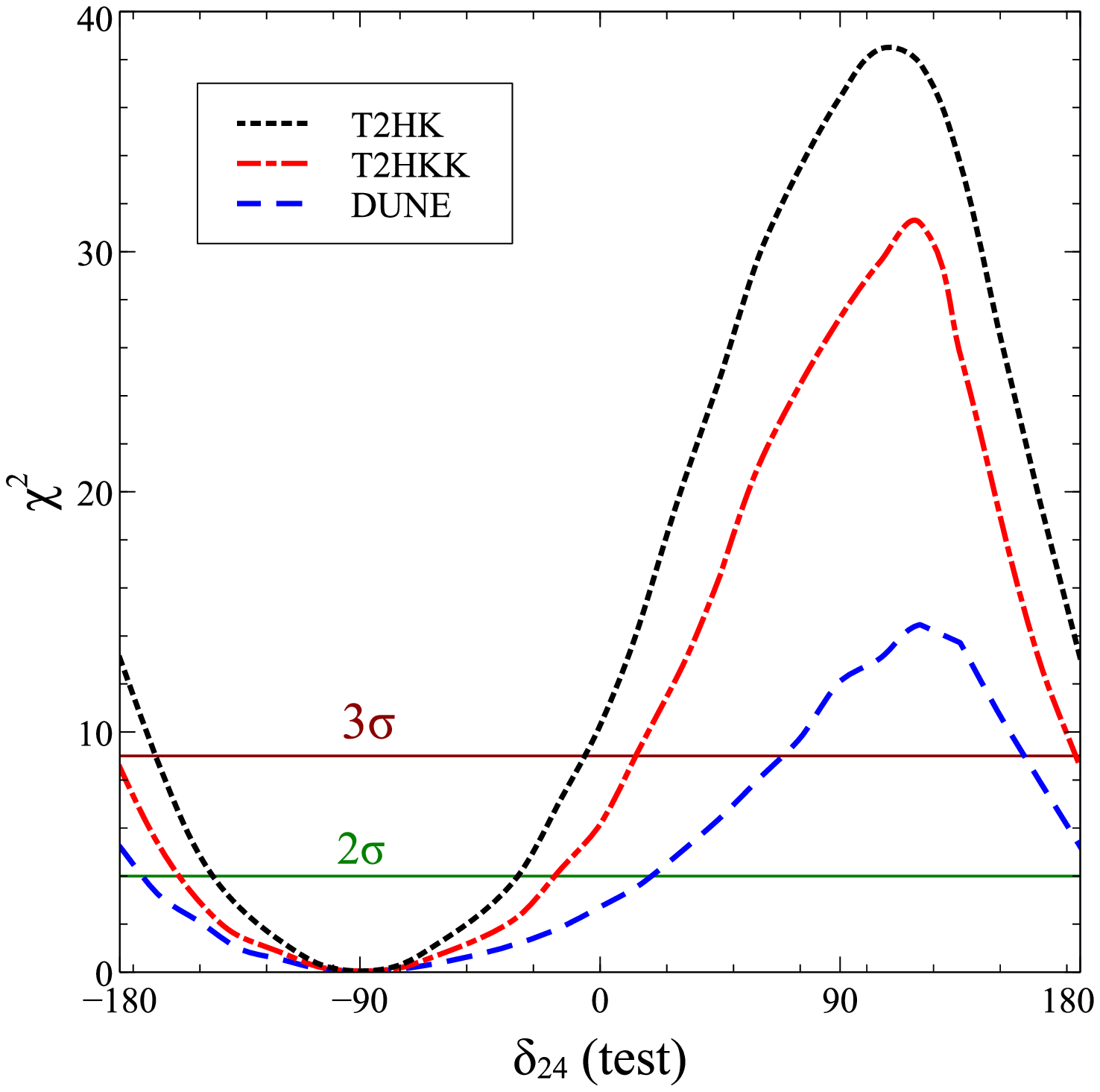}
\includegraphics[width=0.45\textwidth]{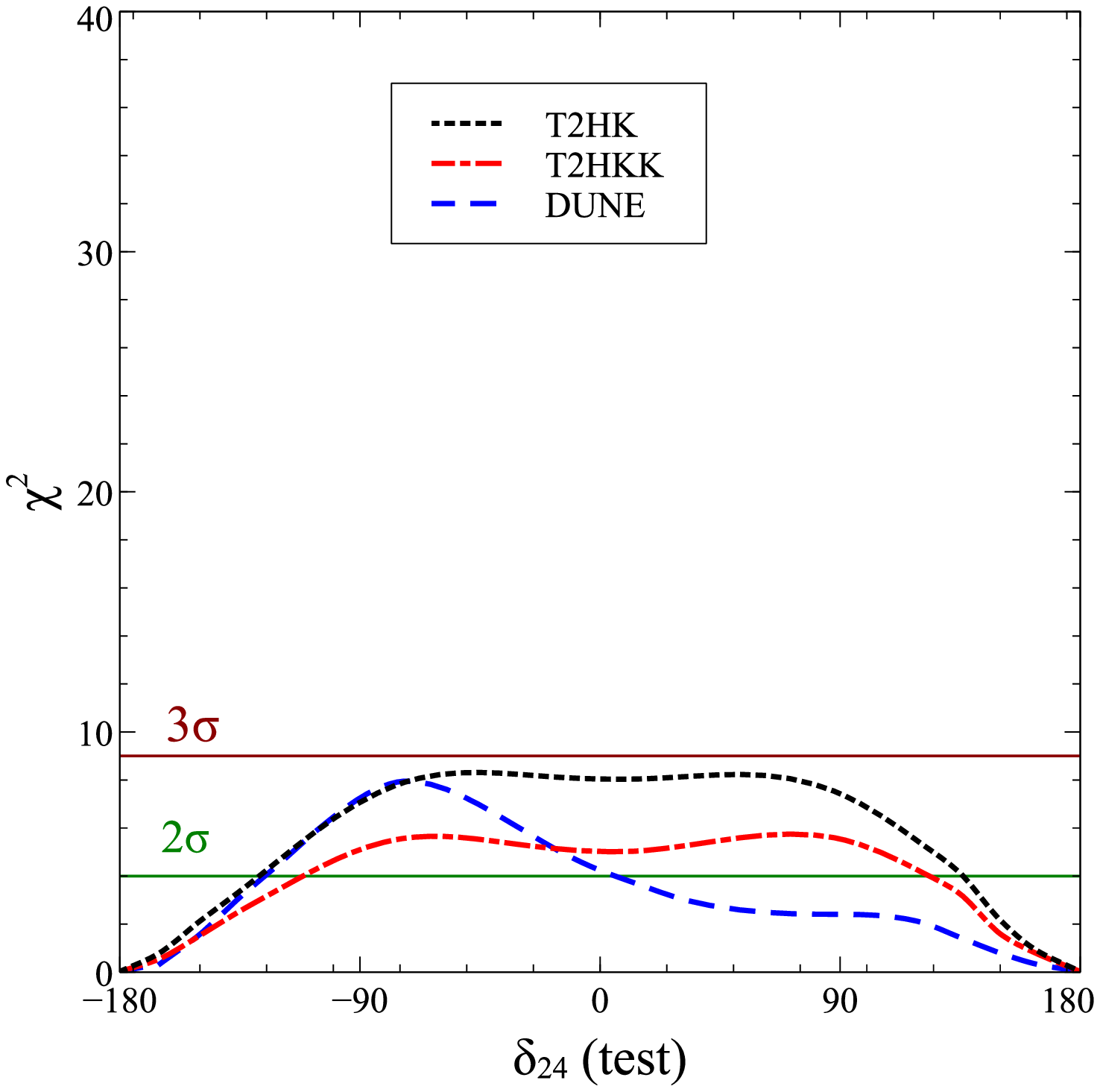}
\caption{\label{1d_phase} The $\chi^{2}$ vs. $\delta_{24}$(test). The black curves are for T2HK, the red curvess are for T2HKK and the blue curves are for DUNE.  The top left panel is for $\delta_{24}$(true)$=0\degree$, the top right panel is for $\delta_{24}$(true)$ = 90\degree$, the bottom left panel is for $\delta_{24}$(true)$ = -90\degree$ and the bottom right panel is for $\delta_{24}$(true)$ = 180 \degree$. }
\end{figure}

\begin{table}[!b]
\begin{center}
\begin{tabular}{|c|c|c|c|c|}
\hline 
\textbf{Exps}  & \pmb{($\delta^{\rm tr}_{24}=0\degree$)}   & \pmb{ ($\delta^{\rm tr}_{24}=90\degree$)} & \pmb{ ($\delta^{\rm tr}_{24}=-90\degree$)} & \pmb{($\delta^{\rm tr}_{24}=180\degree$)}   \tabularnewline 
& \pmb{$\delta_{24}$(test)} &\pmb{$\delta_{24}$(test)}  & \pmb{$\delta_{24}$(test)} & \pmb{$\delta_{24}$(test)}  \tabularnewline
\hline 

\textbf{DUNE}  & $[-175.1\degree, 98.8\degree]$  & [$23.35\degree$, $180\degree$]&[$-172.21\degree,$ $18.49\degree$] &$\delta_{24}\lesssim-126.4\degree,$ $\delta_{24}\gtrsim 6.9\degree$ \tabularnewline
 & &$\delta_{24}\lesssim-175.1\degree$ & & \tabularnewline
\hline 

\textbf{T2HK}  & $[-63.0\degree, 47.9\degree]$& [$30.10\degree$, $147.9\degree$] &[$-145.8\degree$, $-31.60\degree$]& $\delta_{24}\lesssim-128.4\degree,$ $\delta_{24}\gtrsim 136.2\degree$ \tabularnewline
 \hline 
 
\textbf{T2HKK}& $[-94.0\degree, 60.0\degree]$ & [$26.27\degree$, $157.62\degree$] &$[-158.59\degree, -17.51\degree]$ &$\delta_{24}\lesssim-111.8\degree,$ $\delta_{24}\gtrsim 124.3\degree$  \tabularnewline
\hline 

\end{tabular}
\caption{The $2\sigma$ allowed ranges of $\delta_{24}$(test) for the three experiments in the 3+1 scenario. The assumed true value of $\delta_{13}$ is $-90\degree$. We give the allowed ranges of $\delta_{24}$(test) for $\delta_{24}^{\rm tr} {\rm } = 0\degree, 90\degree, -90\degree, 180\degree$. Here $\db^{\rm tr}$ stands for $\db$(true). }
\label{Table1}

\par\end{center}
\end{table}

In this section we discuss the ability of the long baseline experiments to constrain the sterile phases. The ``data'' is generated for the 3+1 scenario for the values of mixing parameters discussed in Section \ref{sec:sim}. In particular, for the sterile neutrino parameters we take the following values: $\Delta m_{41}^2$(true)$=1.7$ eV$^2$, $\theta_{14}$(true)$=8.13\degree$,  $\theta_{24}$(true)$= 7.14\degree$,  $\theta_{34}$(true)$=0\degree$. The true values of the phases $\delta_{24}$ will be taken at some benchmark values and will be mentioned whenever needed. The true values of standard oscillation parameters are taken at their current best-fit values, mentioned in Section \ref{sec:sim}. The $\chi^2$ is marginalised over the relevant oscillation parameters in the 3+1 scenario, as discussed in Section \ref{sec:sim}, where the parameters are allowed to vary within their current $3\sigma$ ranges. Although there are three phases in the 3+1 scenario, the role of the phase $\delta_{34}$ is weak. As was discussed in the previous section, the mixing angle $\theta_{34}$ affects the oscillation probability $P_{\mu e}$ only when matter effects become important. For DUNE and T2HKK earth matter effects are not very strong while for T2HK the effect of earth matter is even weaker. Since the impact of the phase $\delta_{34}$ on  $P_{\mu e}$ is proportional to the mixing angle $\theta_{34}$, the phase $\delta_{34}$ is also less important for $P_{\mu e}$ for the same reason. Moreover, the current global best-fit for the angle $\theta_{34}$ turns out to be zero \citep{Gariazzo:2017fdh}. Therefore, in this work we set $\theta_{34}$(true)$=0\degree$ in the data. As a result the phase $\delta_{34}$ is not expected to be very crucial in our analysis and hence we take  and $\delta_{34}$(true)$=0^\circ$ in the data and show our results only in the $\delta_{13}$ - $\delta_{24}$ plane. We reiterate that the $\chi^2$ is marginalised over the mixing angle $\theta_{34}$ and phase $\delta_{34}$ in the fit, where the mixing angle is allowed to vary between [$0\degree,12\degree$] \citep{Gariazzo:2017fdh}. 

The Fig.~\ref{1d_phase} shows the capability of DUNE, T2HK and T2HKK to measure  the phase $\delta_{24}$. We show the plots of $\chi^2$ as a function of $\delta_{24}$(test) for T2HK (dotted black lines), T2HKK (dash-dotted red lines) and DUNE (dashed blue lines) for $\delta_{24}$(true) of $0\degree$ (top-left panel), $90\degree$ (top-right panel), $-90\degree$ (bottom-left panel) and $180\degree$ (bottom-right panel). The true values of all other parameters are taken as detailed in Section \ref{sec:sim} and the previous paragraph. The $\chi^2$ plot has been marginalised over all relevant parameters as discussed before. The green solid horizontal lines show the $\Delta \chi^2$ corresponding to $2\sigma$ C.L. 
Table~\ref{Table1} shows that T2HK can better constrain the phase $\delta_{24}$ as compared to DUNE, while T2HKK is expected to perform better than DUNE but worse than T2HK.
Note that the sensitivity of DUNE and T2HKK is marginally better for $\delta_{24}$(true)$=90\degree$ than for $\delta_{24}$(true)$=-90\degree$ while the reverse is true in case of T2HK (see Table~\ref{Table1}). 

\begin{figure}{}
\includegraphics[width=0.45\textwidth]{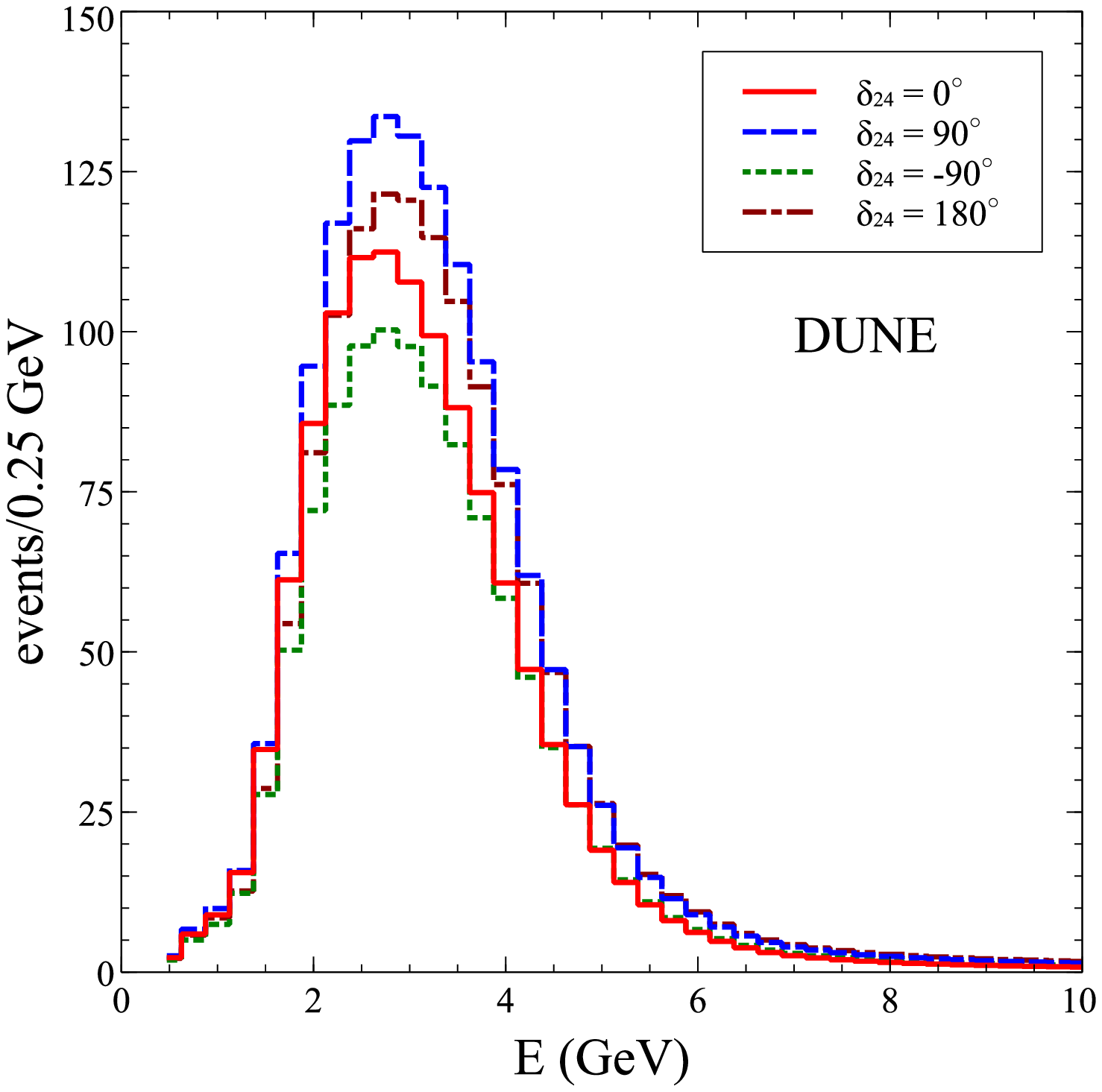}
\includegraphics[width=0.45\textwidth]{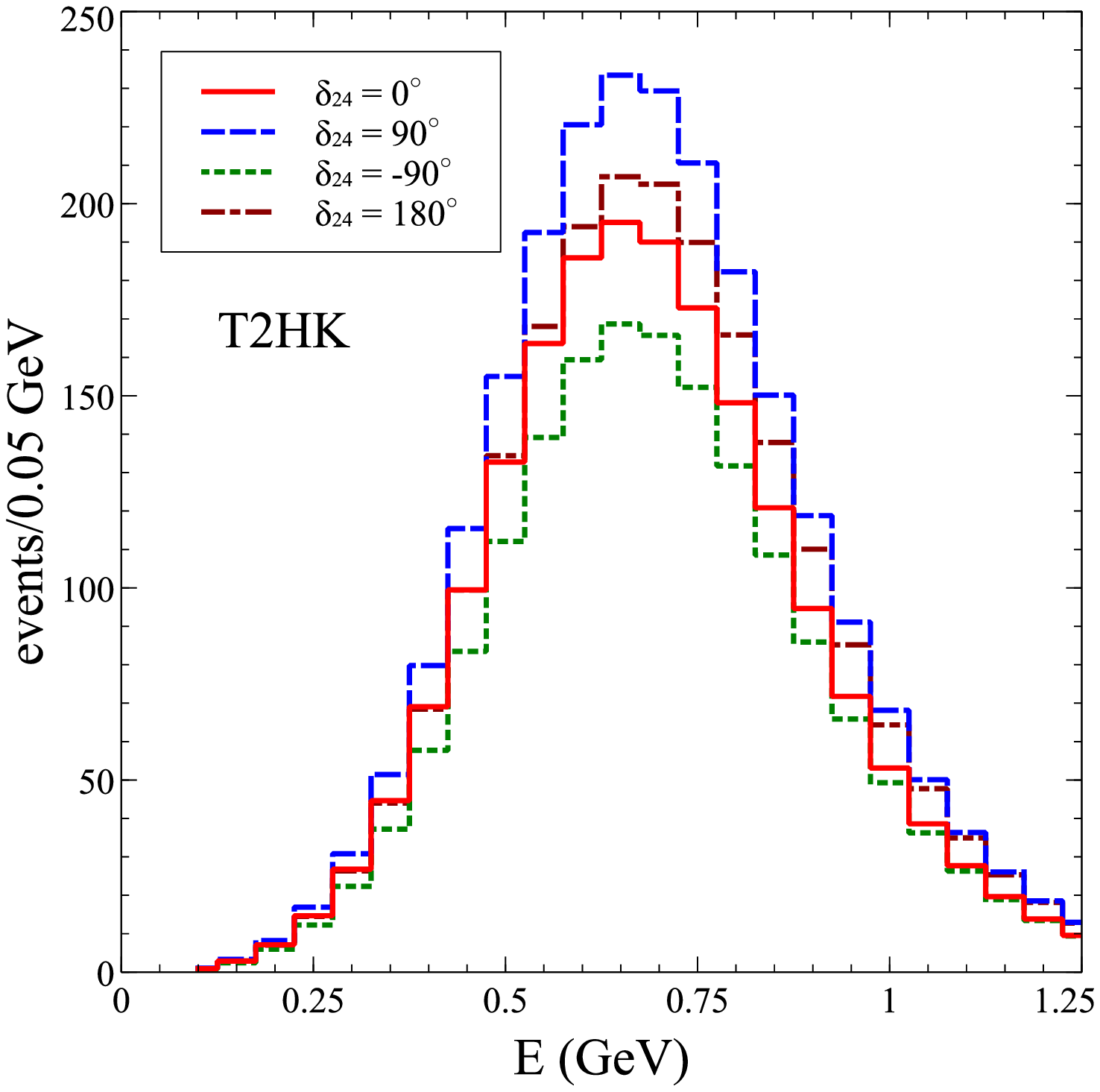}
\includegraphics[width=0.45\textwidth]{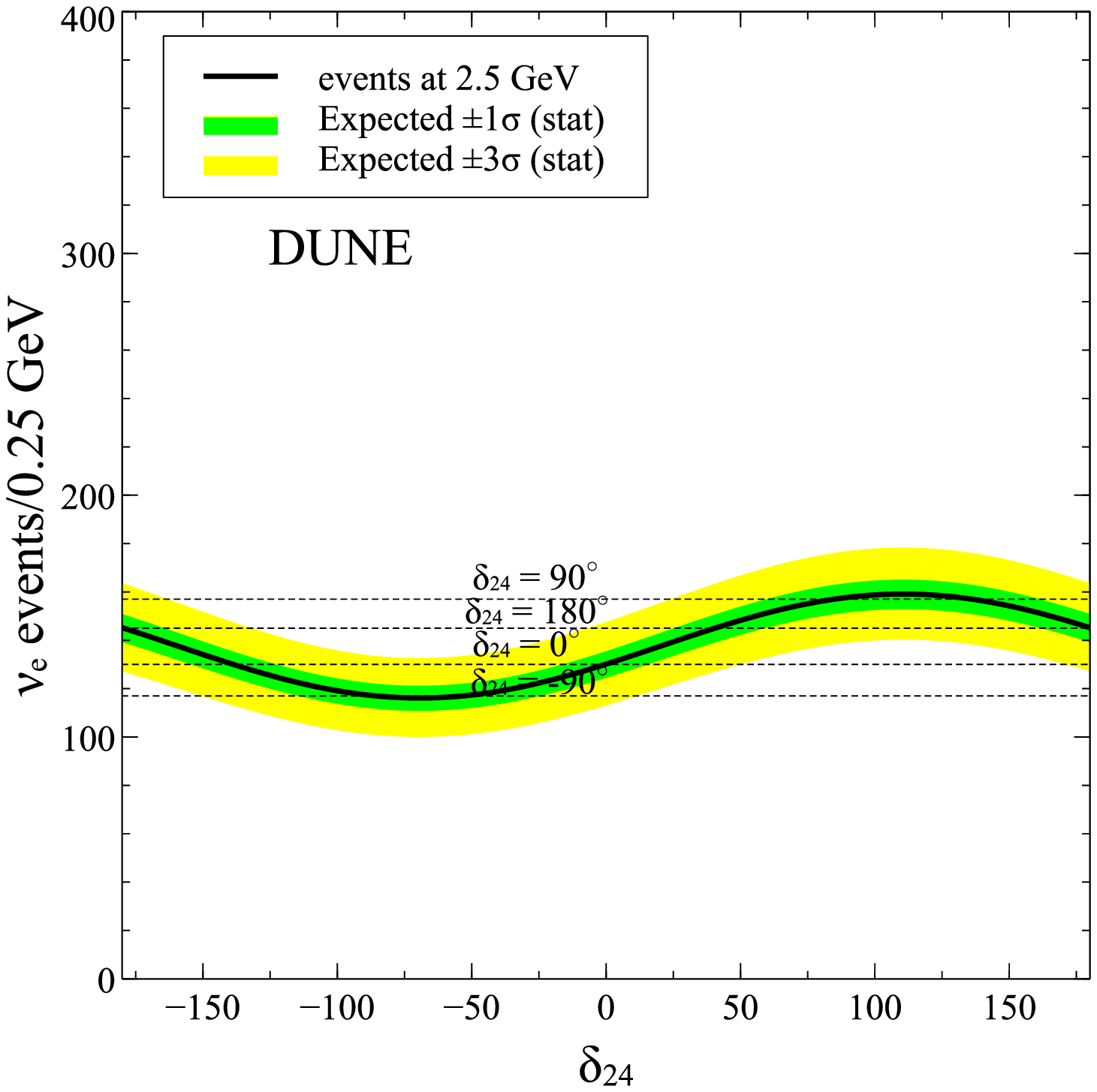}
\includegraphics[width=0.45\textwidth]{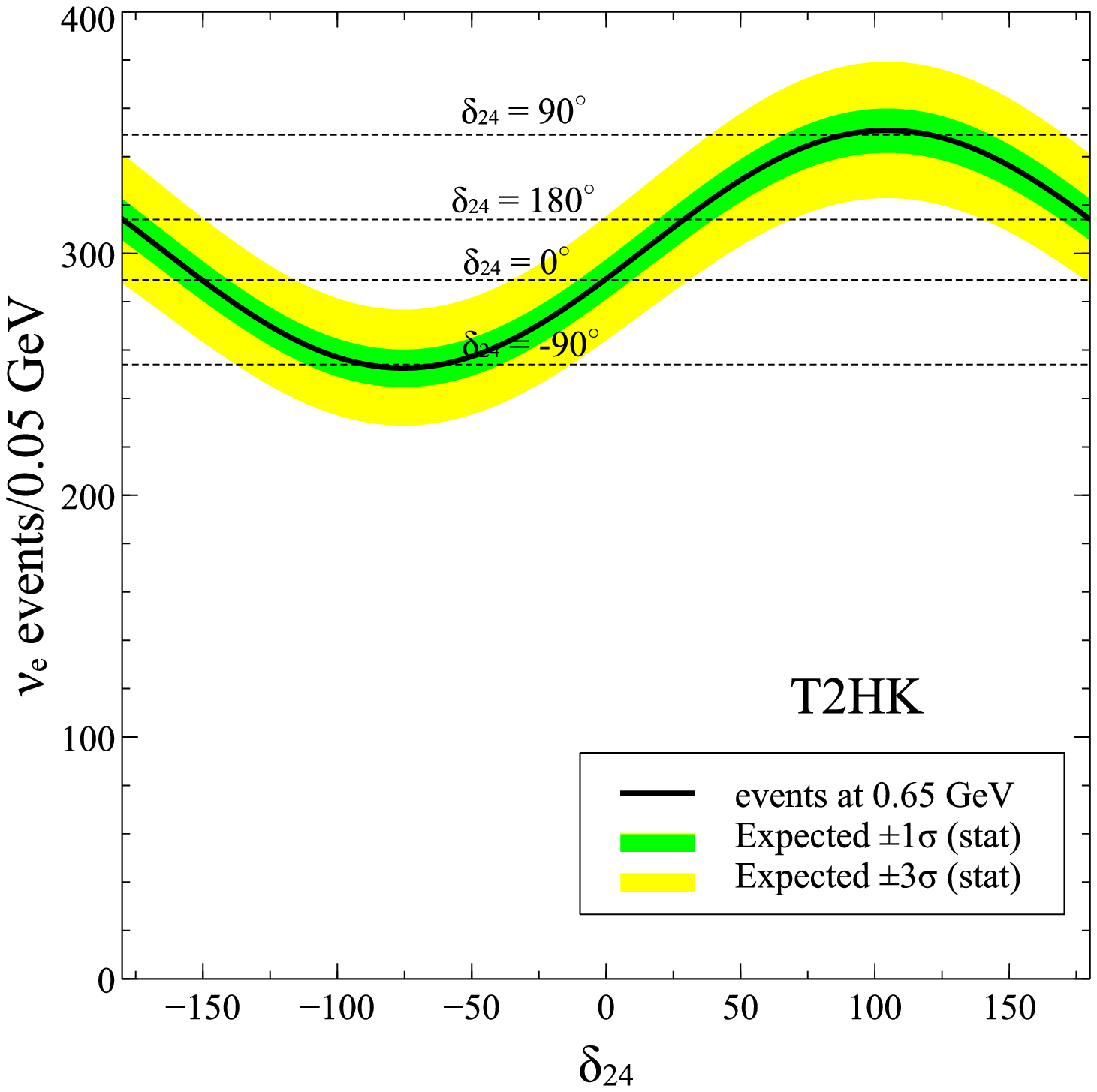}
\caption{\label{event} Top panels show the appearance event spectrum for DUNE (left) and T2HK (right) for different values of $\delta_{24}$ The green lines are for $\delta_{24} = -90\degree$, red lines are for $\delta_{24}=0\degree$, the blue lines are for  $\delta_{24}=90\degree$ and the dark red lines are for $\delta_{24}=180\degree$. The lower panels show the appearance event rates at the oscillation maximum as a function of $\delta_{24}$ for DUNE (left) and T2HK (right). While the black curves give the expected number of events, the green and yellow bands show the 1$\sigma$ and 3$\sigma$ statistical uncertainties.}
\end{figure}

In order to understand why the measurement of $\delta_{24}$ is expected to be better at T2HK than DUNE, we show in Fig.~\ref{event} the expected electron events at DUNE (top left panel) and T2HK (top right panel). The four lines in each panel show the expected events for four values of $\delta_{24}=0\degree$ (solid red lines), $90\degree$ (dashed blue lines), $-90\degree$ (dotted green lines) and $180\degree$ (dash-dotted dark red lines). The upper panels of the figure reveal that the two experiments behave in almost the same way as far as the dependence of the probability $P_{\mu e}$ to $\delta_{24}$ is concerned. However, there is a clear difference between the two when it comes to the overall statistics. T2HK expects to see nearly 14 times more events than DUNE due to its bigger detector size. Hence the corresponding $\chi^2$ for T2HK is also expected to be higher. Of course the systematic uncertainty for DUNE is considerably less than for T2HK and that compensates the effect of the lower statistics, however, the effect of statistic shows up in a non-trivial way for $\delta_{24}$ measurement at the long baseline experiments and T2HK with its bigger detector emerges as a better option in this regard.

The lower panels of  Fig.~\ref{event} show the event rate at the oscillation maximum for DUNE (lower left panel) and T2HK (lower right panel) as a function of $\delta_{24}$. The black solid curves show the expected number of events whereas the green and yellow bands show the 1$\sigma$ and 3$\sigma$ statistical deviation.  The black short-dashed straight lines show the event rate at oscillation maxima for the four benchmark values of $\delta_{24}=0\degree,~180\degree,~90\degree$ and $-90\degree$. The Fig.~\ref{1d_phase} had revealed that the $\chi^{2}$ corresponding to $\delta_{24}$(true)$=0\degree$ and  $\delta_{24}$(true)$=180\degree$ are much lower compared to that for $\delta_{24}$(true)$=\pm90\degree$. This can be understood from the lower panels of Fig.~\ref{event} as follows. Fig.~\ref{event}  shows that the predicted number of events at oscillation maximum for $\delta_{24}$(true)$=0\degree$ and $180\degree$ lie between the predicted events for $\delta_{24}$(true)$=\pm 90\degree$. Therefore, for the cases where data is generated for $\delta_{24}$(true)$=0\degree$ and $180\degree$, it is easier for other $\delta_{24}$  values to fit the data and give a smaller $\chi^2$. However, data corresponding to $\delta_{24}$(true)$=\pm 90\degree$ takes a more extreme value and the difference between the data and fit for other values of $\delta_{24}$ for these cases becomes larger, giving larger $\chi^2$.

Another interesting feature visible in the event plots in Fig.~\ref{event} is that the maxima and minima of the events are not at $\delta_{24} = 90\degree$ or $-90\degree$. Rather they are slightly shifted towards the right. For the same reason the $\chi^{2}$ plots in Fig.~\ref{1d_phase} are also  asymmetric about the true value of $\delta_{24}$. One can explain this using Eq.~(\ref{eq:pmue}).  By inspecting the probability one can see that the
correlation between $\ms$ and $\delta_{24}$ is negligible. Also, we have taken  $\delta_{13} =-90\degree$ everywhere. Hence, for $\ms=0$ and  $\delta_{13} =-90\degree$, the Eq.~(\ref{eq:pmue}) can be rearranged as,
\begin{equation}
P_{\mu e} = A + B\cos2\theta_{13}\sin\delta_{24}-\frac{1}{2}B\cos\delta_{24}
\,,
\label{eq:pmue_simp}
\end{equation}
where A and B are independent of $\delta_{24}$. In the absence of the last term, the probability would be a sine function shifted by the constant A. However, the presence of the cosine term shifts the curve and the shift is towards right because of the minus sign in front of the cosine term. In particular, the extrema of the probability in Eq.~(\ref{eq:pmue_simp}) is given by the condition,
\be
\cos\delta_{24}=-\frac{1}{2}\sin\delta_{24}
\,,
\ee
which corresponds to minimum at $\delta_{24}=-63.4^\circ$ and maximum at $\delta_{24}=116.6^\circ$. This agrees very well with the event plots in the lower panels of Fig.~\ref{event} which is obtained using the exact numerical probability. 


\begin{figure}{}
\centering
\includegraphics[width=0.45\textwidth]{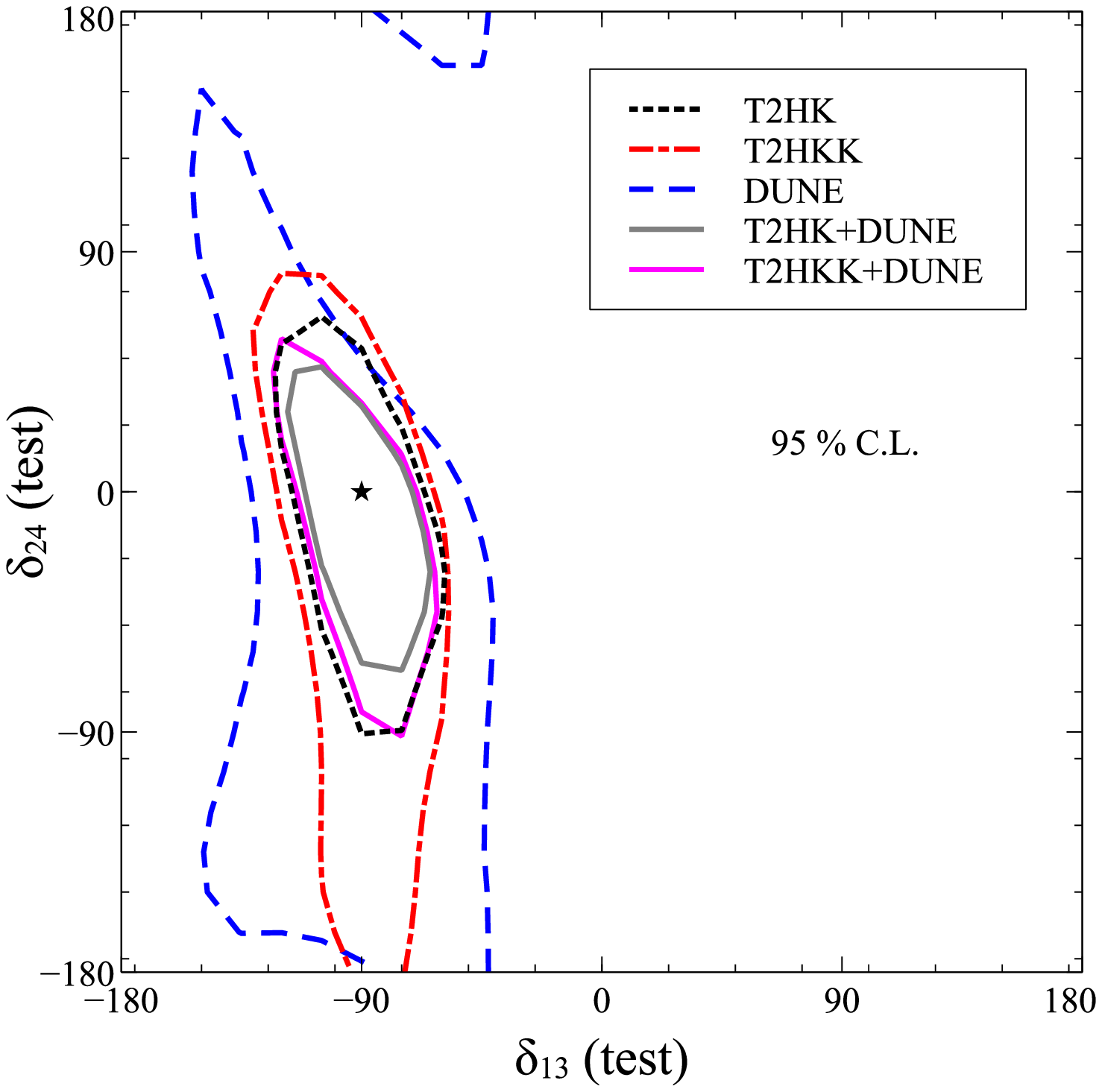}
\includegraphics[width=0.45\textwidth]{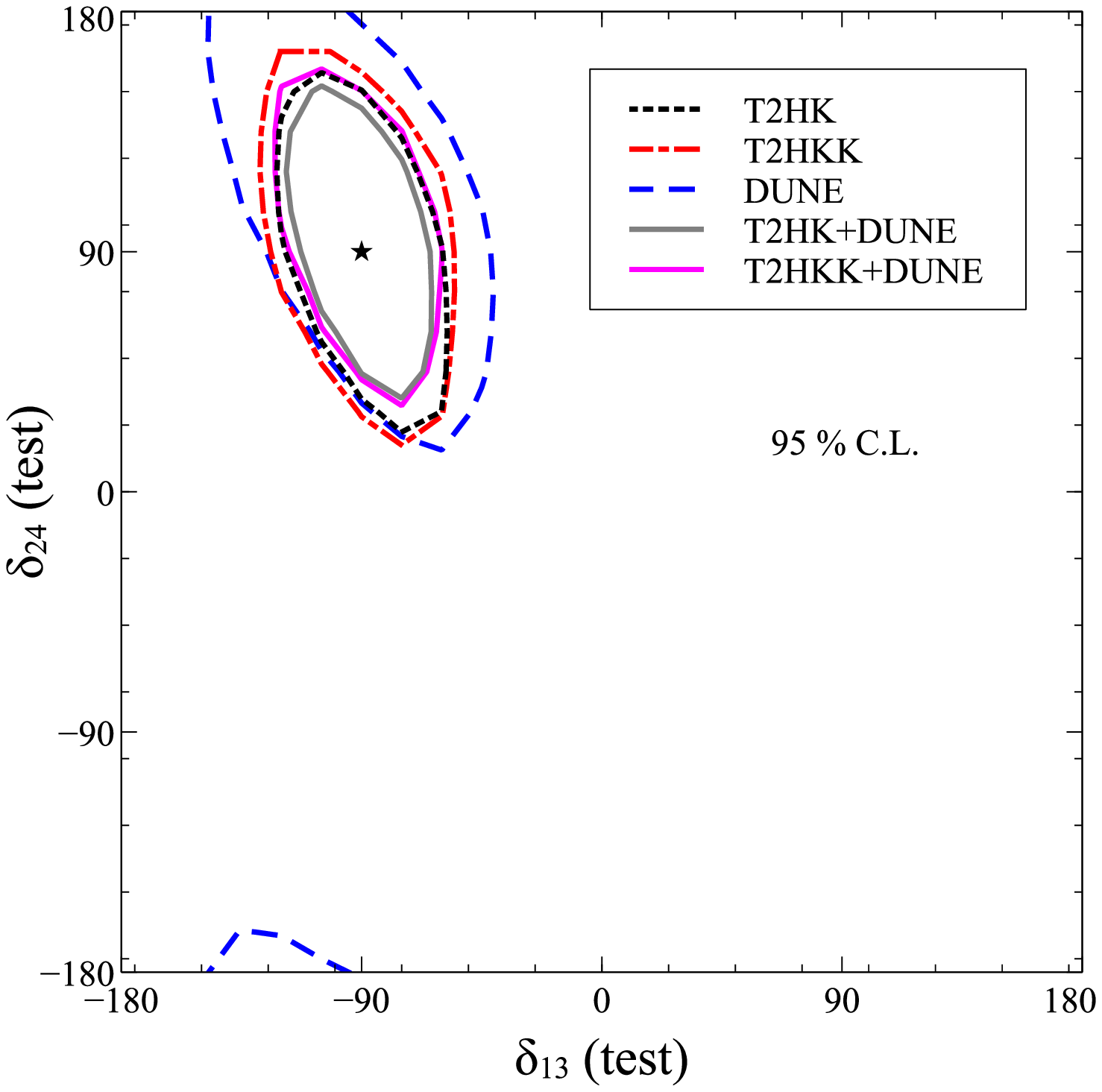}
\includegraphics[width=0.45\textwidth]{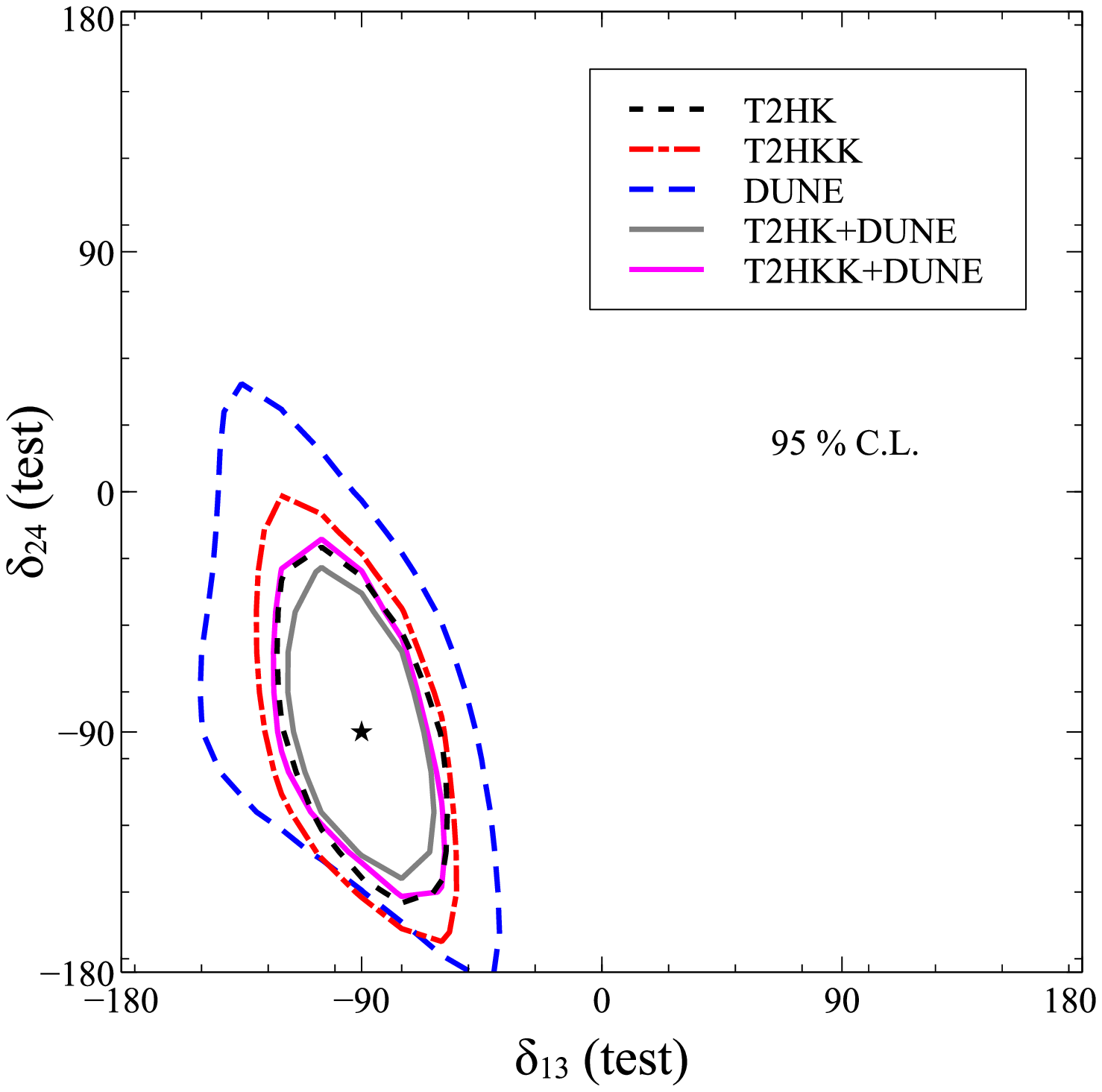}
\includegraphics[width=0.45\textwidth]{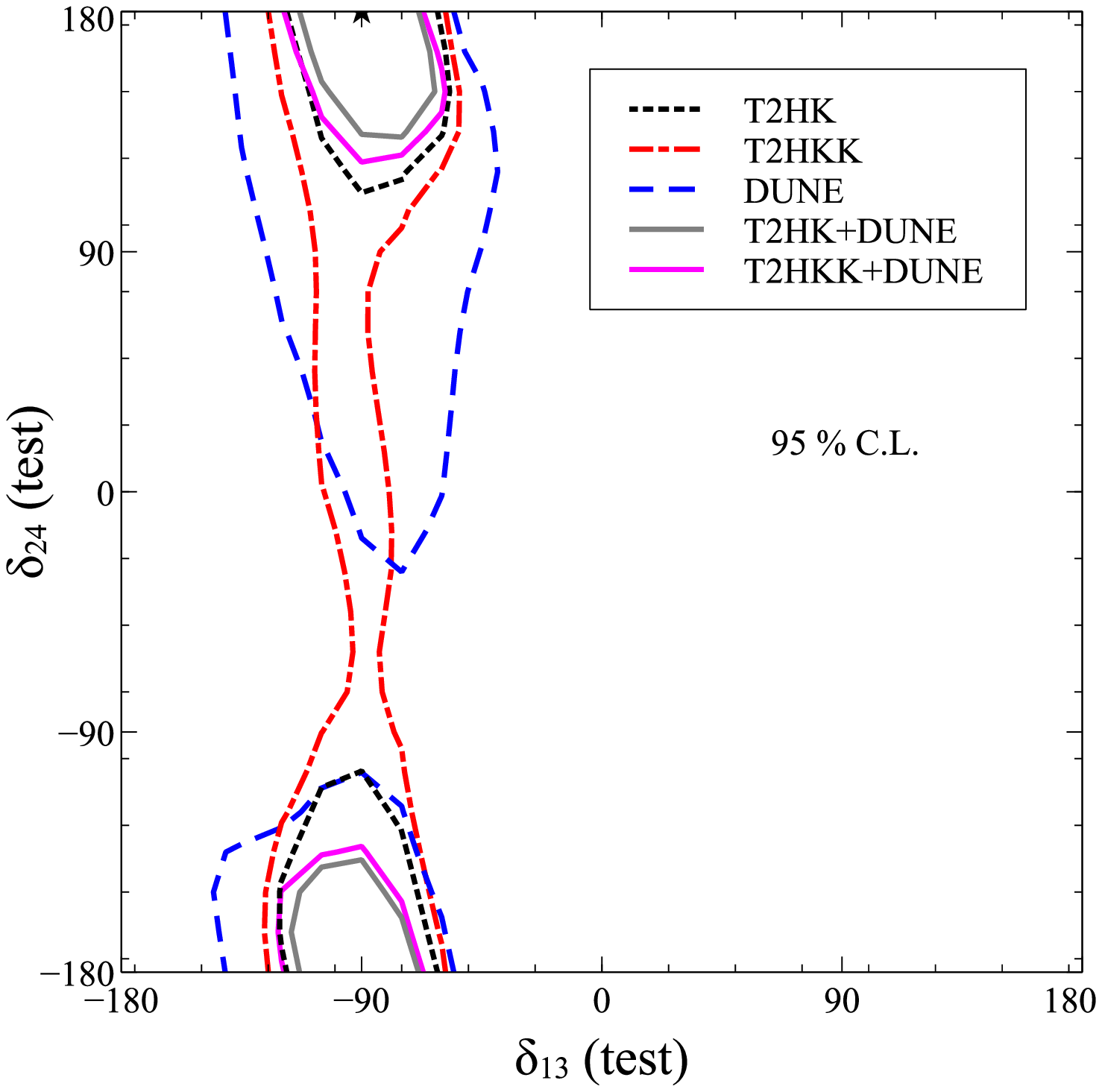}
\caption{\label{phasen0}  The expected 95 \% C.L. contours in $\delta_{13}$(test) vs $\delta_{24}$(test), where 95 \% C.L. is defined as $\Delta\chi^2=5.99$ for 2 parameters. The stars show the value of $\delta_{13}$(true) and $\delta_{24}$(true) taken in the data. The top left (right) panel is for $\delta_{24}=0\degree$ ($\delta_{24} =90\degree $) and the bottom left (right) panel is for $\delta_{24} = -90\degree$ ($\delta_{24} = 180\degree$). 
The black dotted curve is for T2HK, the red dash-dotted curve is for T2HKK, the blue dashed curve is for DUNE, the grey solid curve is for DUNE+T2HK and the magenta solid curve is for DUNE + T2HKK.}
\end{figure}

We next show in Fig.~\ref{phasen0} the expected 95 \% C.L. allowed areas in the $\delta_{13}$(test)$- \delta_{24}$(test) plane, expected to be measured by the next generation long-baseline experiments T2HK (or T2HKK) and DUNE, and by combining them. The four panels of Fig.~\ref{phasen0} have been generated for four different choices of $\delta_{24}$(true). The value of $\delta_{13}$(true)$=-90\degree$ in all the panels. In each panel the benchmark point where the data is generated is shown by the black star. The four panels correspond to $\delta_{24}$(true)$ = 0\degree$ (top left), $90\degree$ (top right), $-90\degree$ (bottom left) and $180\degree$ (bottom right). In all the four cases we have considered 3+1 scenario both in the `data' and in the `fit' or  `theory'. The $\chi^2$ thus generated is then marginalised over the sterile mixing angles $\theta_{14}$, $\theta_{24}$, $\theta_{34}$ and $\delta_{34}$, as discussed before. The black dotted, red dash-dotted and blue dashed contours are for T2HK, T2HKK and DUNE, respectively, while the grey and magenta solid contours are for DUNE+T2HK and DUNE+T2HKK. As in Fig.~\ref{1d_phase} we note that T2HK can constrain the phase $\delta_{24}$ much better than DUNE, while T2HKK performs better than DUNE but worse than T2HK. We also see, as before, that for DUNE the precision on $\delta_{24}$ is expected to be better for $\delta_{24}$(true)$=\pm 90\degree$ compared to when $\delta_{24}$(true)$=0\degree$ or $180\degree$. For T2HK this dependence of precision  on $\delta_{24}$ measurement on $\delta_{24}$(true) is less pronounced. The effect of $\theta_{34}$ on the measurement of $\delta_{24}$ is also minimal. Finally, note that there is an anti-correlation between $\da$ and $\db$. This comes from the term $P_4(\da+\db)$ of Eq.~(\ref{eq:pmue}). 

The Fig.~\ref{phasen0} also shows how the measurements of $\delta_{24}$ and $\delta_{13}$ improve as we combine DUNE with either T2HK or T2HKK. We see that combining DUNE with T2HKK improves the precision considerably, with the combined precision of DUNE and T2HKK becoming slightly better than the precision expected from T2HK alone. Combining DUNE with T2HK improves the precision even further, albeit only marginally, since T2HK alone can measure the phases rather precisely.

%

\begin{figure}{}
\includegraphics[width=0.32\textwidth]{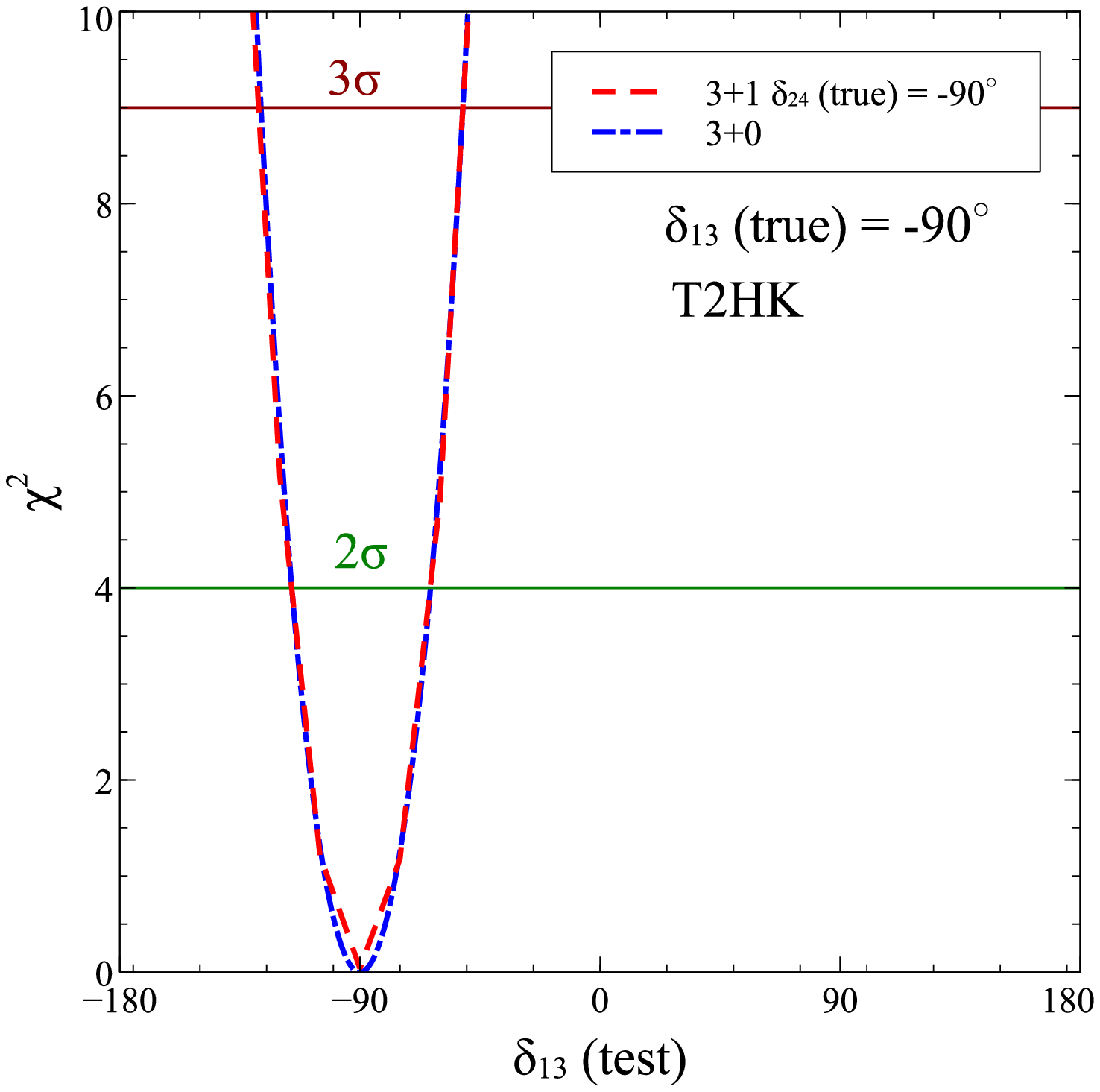}
\includegraphics[width=0.32\textwidth]{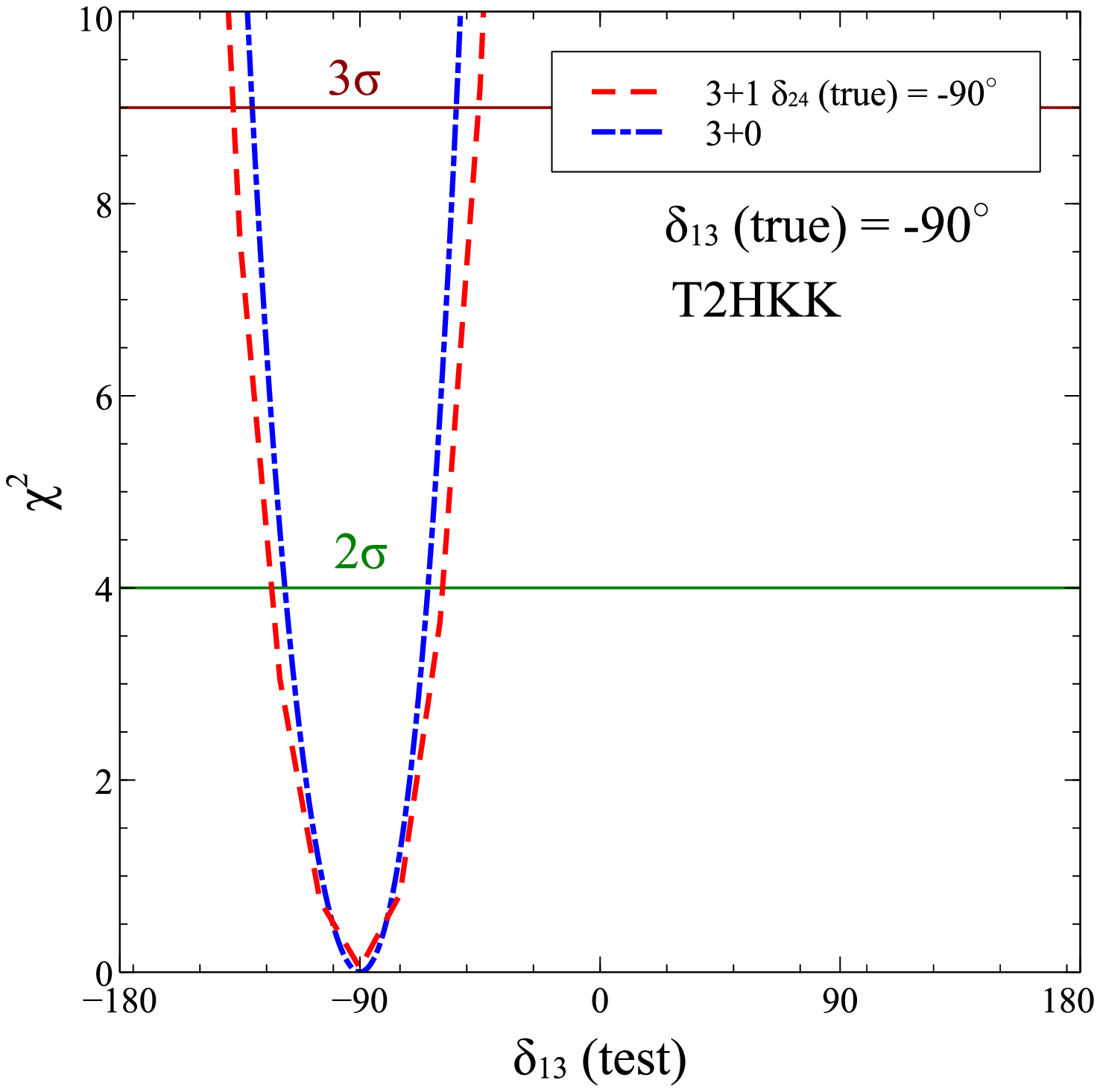}
\includegraphics[width=0.32\textwidth]{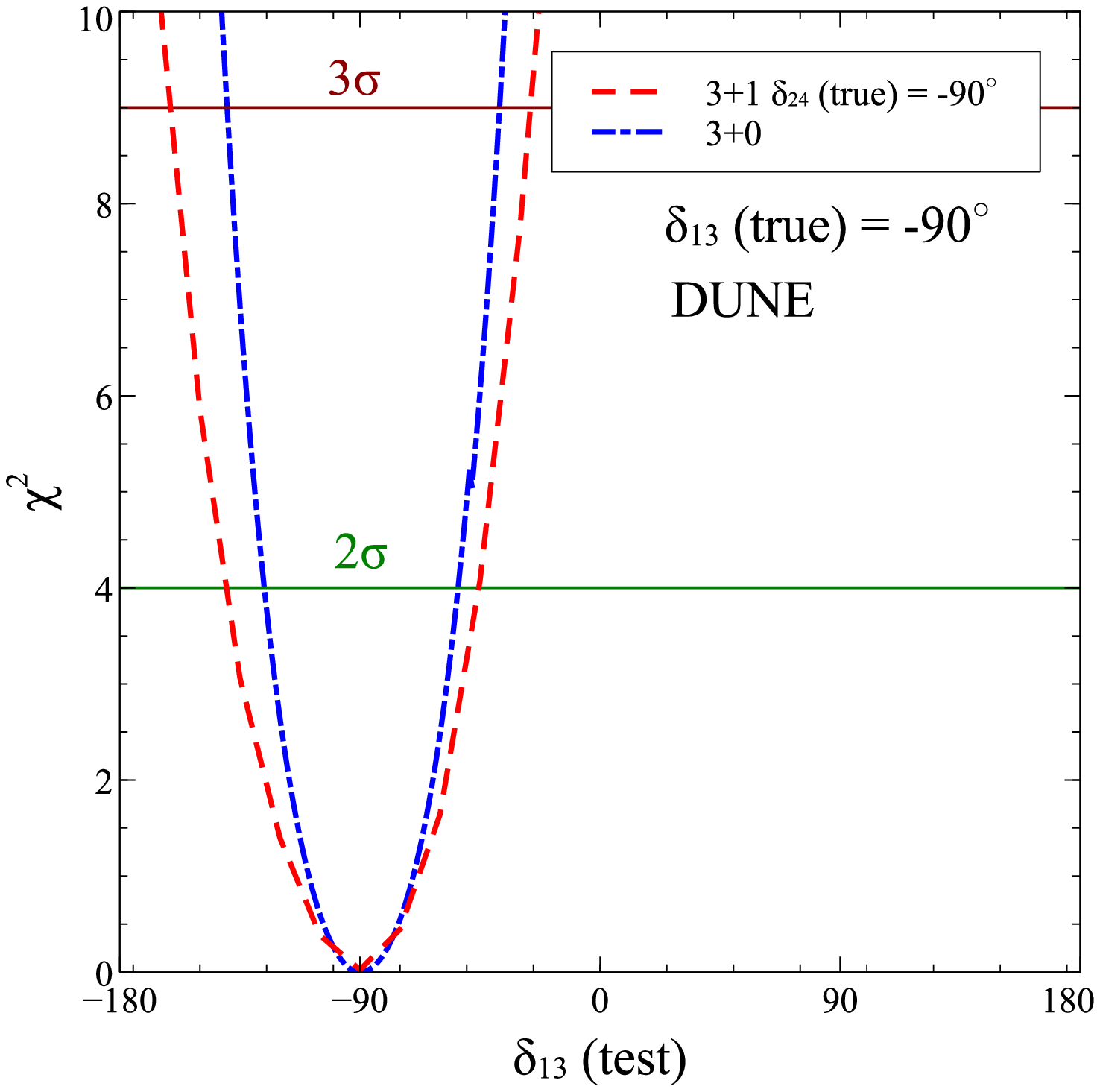}
\caption{\label{del13-2} The expected precision on $\da$  for the 3+0 and 3+1 scenarios. The left panel is for T2HK, the middle panel is for T2HKK and the right panel is for DUNE. The blue dash-dotted curves are for the 3+0 case and the red dashed curves are for 3+1 case in both theory and data. The curves are $\db$(true)$ = -90\degree$. }
\end{figure}

The question on how the measurement of the standard CP phase $\delta_{13}$ gets affected by the sterile mixing angle phases in the 3+1 scenario is another pertinent question that one can ask. The Fig.~\ref{del13-2} shows how the expected precision on $\da$ changes in presence of sterile neutrinos. The left panel is for T2HK, middle panel is for T2HKK and right panel is for DUNE. The blue dashed curves are for the standard 3+0 case with no sterile neutrinos while the red dash-dotted curves are for the 3+1 case with $\db = -90\degree$ in data. The other standard and sterile neutrino oscillation parameters are taken in data as described above and the fit performed as before. The Fig.~\ref{del13-2} shows that the expected precision on $\da$ worsens when the sterile neutrino is present. From Eq.~(\ref{eq:p4}) one can see that there is an anti-correlation between $\da$ and $\db$ which makes the $\da$ precision worse. For DUNE the effect is more compared to T2HK and T2HKK. For T2HK the $\da$ measurement is seen to be nearly unaffected. DUNE measures $\db$ worse than T2HK and T2HKK and hence the corresponding measurement of $\da$ worsens due to the anti-correlation mentioned above. Table \ref{Table2} summarises the expected precision on $\da$ for the 3+0 and 3+1 scenario for four benchmark values of $\db$(true). We see that DUNE's measurement of $\da$ gets affected for all $\db$ while effect on T2HK's measurement of $\da$ is negligible. 
 
 \begin{table}[!h]
\begin{center}
\begin{tabular}{|c|c|c|c|c|c|}
\hline 
\textbf{Exps} & \pmb{3+0($\delta^{\rm tr}_{13}=-90\degree$)} & \pmb{3+1 ($\delta^{\rm tr}_{24}=0\degree$)}   & \pmb{3+1 ($\delta^{\rm tr}_{24}=90\degree$)} & \pmb{3+1 ($\delta^{\rm tr}_{24}=-90\degree$)} & \pmb{3+1 ($\delta^{\rm tr}_{24}=180\degree$)}   \tabularnewline 
&\pmb{$\delta_{13}$(test)} & \pmb{$\delta_{13}$(test)} &\pmb{$\delta_{13}$(test)}  & \pmb{$\delta_{13}$(test)} & \pmb{$\delta_{13}$(test)}  \tabularnewline
\hline 
\textbf{DUNE} & $[-125.3\degree, -53.6\degree]$ & $[-143.3\degree, -48.8\degree]$  & $[-139.1\degree, -48.2\degree]$&$[-139.0\degree, -46.2\degree]$ &$[-137.6\degree, -45.2\degree]$\tabularnewline
 
\hline 
\textbf{T2HK} &$[-115.3\degree, -64.2\degree]$ & $[-116.9\degree, -66.6\degree]$& $[-114.0\degree, -65.5\degree]$ &$[-113.2\degree, -65.0\degree]$ & $[-113.9\degree, -63.5\degree]$ \tabularnewline

\hline 
\textbf{T2HKK} &$[-117.6\degree, -65.1\degree]$ & $[-123.5\degree, -63.2\degree]$ & $[-121.1\degree, -60.0\degree]$ &$[-121.3\degree, -58.1\degree]$ &$[-121.1\degree, -57.3\degree]$ \tabularnewline

\hline 
\end{tabular}
\caption{The $2\sigma$ allowed $\delta_{13}$(test) ranges for the three experiments both in 3+0 and 3+1 scenario. In both the scenarios, assumed true value of $\delta_{13}$ is $-90\degree$ while in 3+1 case, we give the allowed ranges of $\delta_{13}$(test) for $\delta^{\rm tr}_{24} = 0\degree, 90\degree, -90\degree, 180\degree$. Here $\delta^{\rm tr}$ stands for $\delta$(true).}
\label{Table2}

\par\end{center}
\end{table}

\section{Measurement of the mixing angles\label{sec:mixangle}}

Prospects of measuring the sterile neutrino mixing angles at long-baseline experiments DUNE \citep{Berryman:2015nua} and T2HK \citep{Kelly:2017kch} has been studied before. Here we study how well the sterile mixing can be constrained by combining data from these experiments. We also present the sensitivity of the individual experiment DUNE, T2HK and T2HKK. Here we consider two complementary approaches. We first assume that sterile neutrino mixing does exist (as in the last section) and see how precisely the data from long-baseline experiments can measure and constrain the angles $\theta_{14}$ and $\theta_{24}$\footnote{We do not study the mixing angle $\theta_{34}$ in this work. As discussed before, this affects $P_{\mu e}$ and $P_{\mu\mu}$ only mildly through matter effects. To constrain this angle, we need to consider the neutral current data, which has been done in \citep{Adamson:2017zcg, Gandhi:2017vzo, Coloma:2017ptb}.}. We next consider the alternate situation where the active-sterile oscillations do not really exist and then we see how well DUNE, T2HK and T2HKK, as well as their combination, could put upper bounds on the sterile neutrino mixing angles $\theta_{14}$ and $\theta_{24}$. 

\subsection{Measuring the Sterile Mixing Angles when 3+1 is True \label{sec:withsterile}}

\begin{figure}
\includegraphics[width=0.45\textwidth]{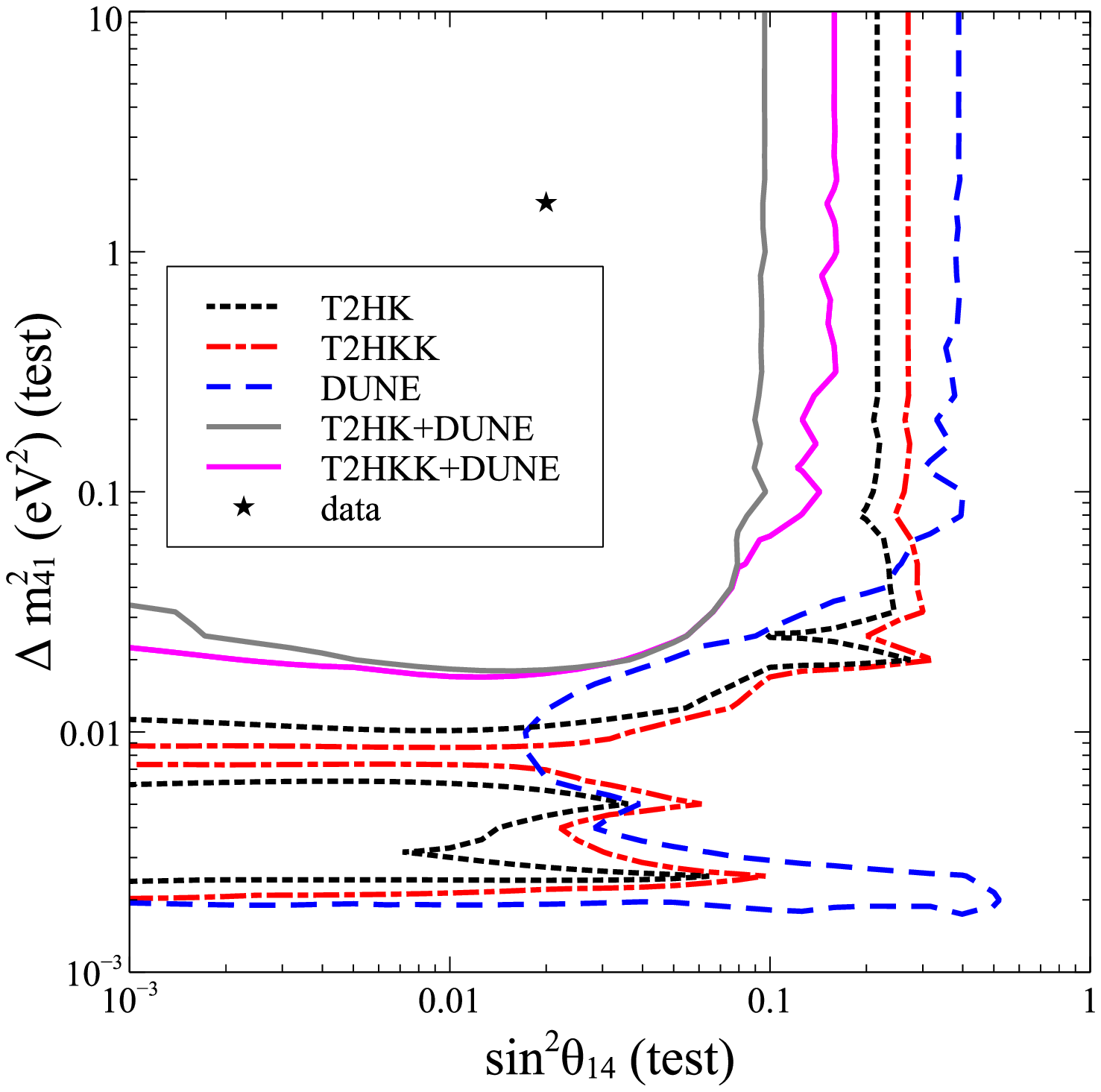}
\includegraphics[width=0.45\textwidth]{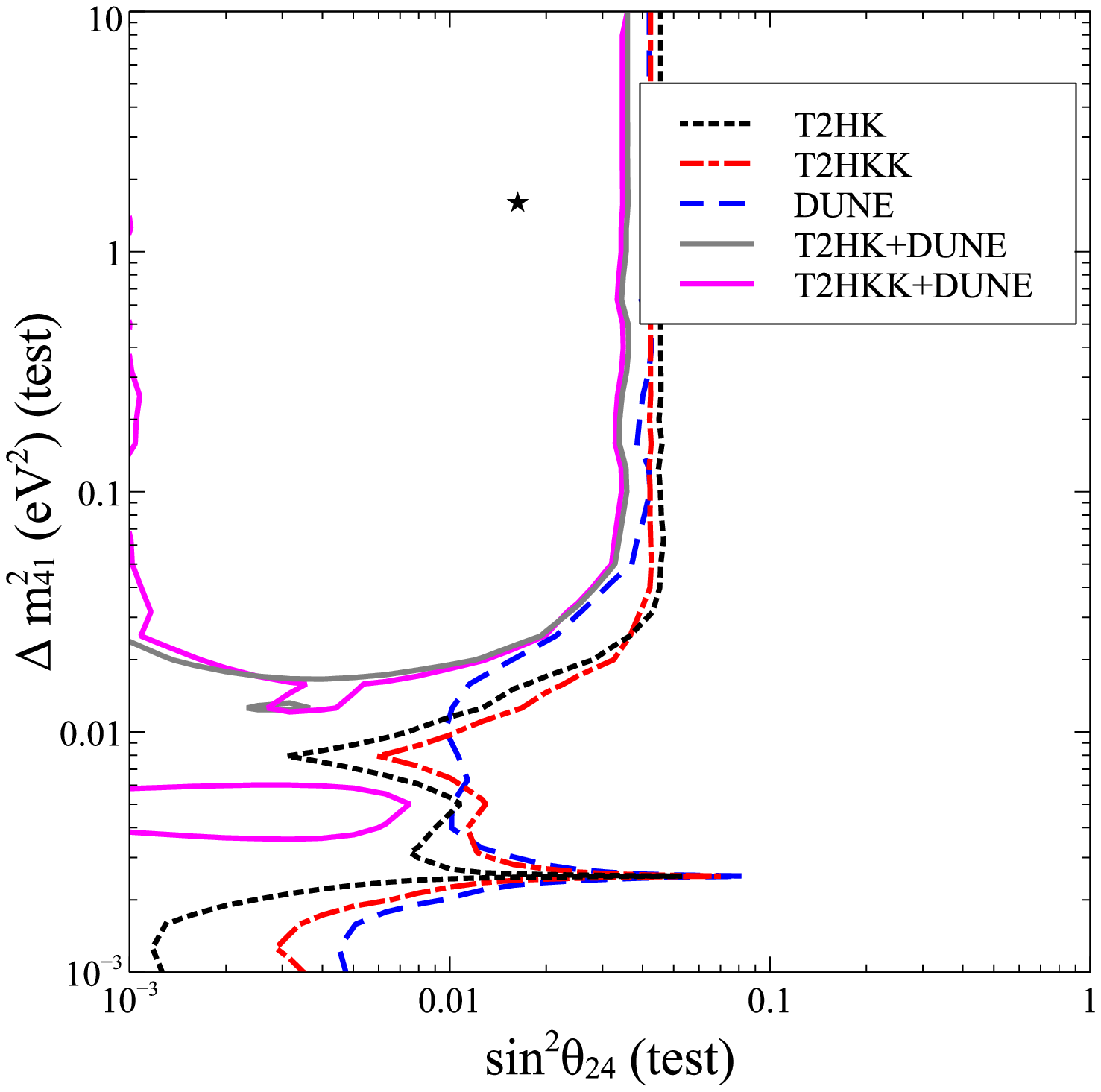}
\caption{\label{dm41n0}
The left panel shows the expected 95 \% C.L contours in the $\sin^{2}\theta_{14}$(test)-$\Delta m^{2}_{41}$(test) plane, while the right panel shows the 95 \% C.L. contours in the $\sin^{2}\theta_{24}$(test)-$\Delta m^{2}_{41}$(test) plane. The colour code is same as Fig.~\ref{phasen0}.} 
\end{figure}

In this subsection we assume that the 3+1 scenario is indeed true in nature and the mixing angles $\theta_{14}$ and $\theta_{24}$ are indeed non-zero. We perform a $\chi^2$ analysis with prospective data generated in the 3+1 scenario and fitted within the 3+1 scenario and give expected allowed C.L. regions in the sterile neutrino parameter spaces. As before, we take the true  sterile oscillation parameters at the following benchmark values: $\Delta m^{2}_{41}$(true)$ = 1.7 $ eV$^{2}$, $\theta_{14}$(true)$ = 8.13\degree$, $\theta_{24}$(true)$ = 7.14\degree$,  $\theta_{34}$(true)$ = 0\degree$ which are consistent with \citep{Gariazzo:2017fdh}. The standard oscillation parameters are taken and treated as discussed before. The $\chi^2$ is marginalised over all relevant parameters in the fit and no Gaussian priors are included. 

In Fig.~\ref{dm41n0}, we show the contours in $\sin^{2}\theta_{14}$(test)-$\Delta m^{2}_{41}$(test) plane (left panel) and $\sin^{2}\theta_{24}$(test) - $\Delta m^{2}_{41}$(test) plane (right panel). The colour code is same as Fig.~\ref{phasen0}. The point where the data is generated is shown by the black star in the two panels. The results show that in both panels, T2HK gives better results than both DUNE and T2HKK. Again, T2HKK is better than DUNE. 
Combining T2HK/T2HKK and DUNE experiments improves the results and the expected allowed ranges for the sterile neutrino mixing parameters shrink. 
The precision expected from the combined data from T2HK and DUNE is nearly the same as that from T2HKK and DUNE, which for the former is only marginally better.

\begin{figure}
\includegraphics[width=0.5\textwidth]{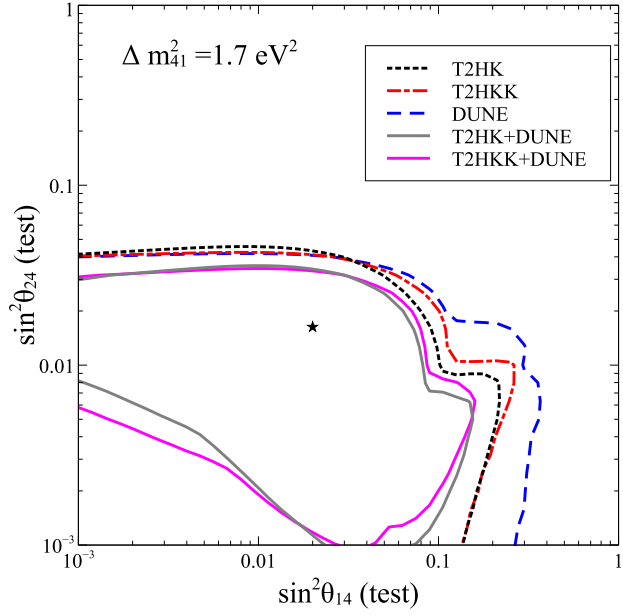}
\caption{\label{th14th24n0}The expected 95 \% C.L contours in $\sin^{2}\theta_{24}$(test)-$\sin^{2}\theta_{14}$(test) plane. The colour code is same as Fig.~\ref{phasen0}.}
\end{figure}

Fig.~\ref{th14th24n0} shows the contours in the $\sin^{2}\theta_{14}$(test)-$\sin^{2}\theta_{24}$(test). The colour code for the different data sets considered is the same as Fig.~\ref{phasen0}. We note that the expected upper limit on $\sin^{2}\theta_{24}$ is the same for 
all the three individual experiments for $0.001< \sin^{2}\theta_{14} < 0.03$. Also, the expected upper bound on $\sin^2\theta_{14}$ is seen to be better for T2HK than DUNE. Combining the experiments can improve the measurement of the sterile neutrino mixing angles as seen from the solid contours in Fig.~\ref{th14th24n0}. In particular, we now see a lower bound on  $\sin^{2}\theta_{24}$. 

\subsection{Excluding the Sterile Hypothesis when 3+1 is Not True\label{sec:nosterile}}

\begin{figure}
\includegraphics[width=0.45\textwidth]{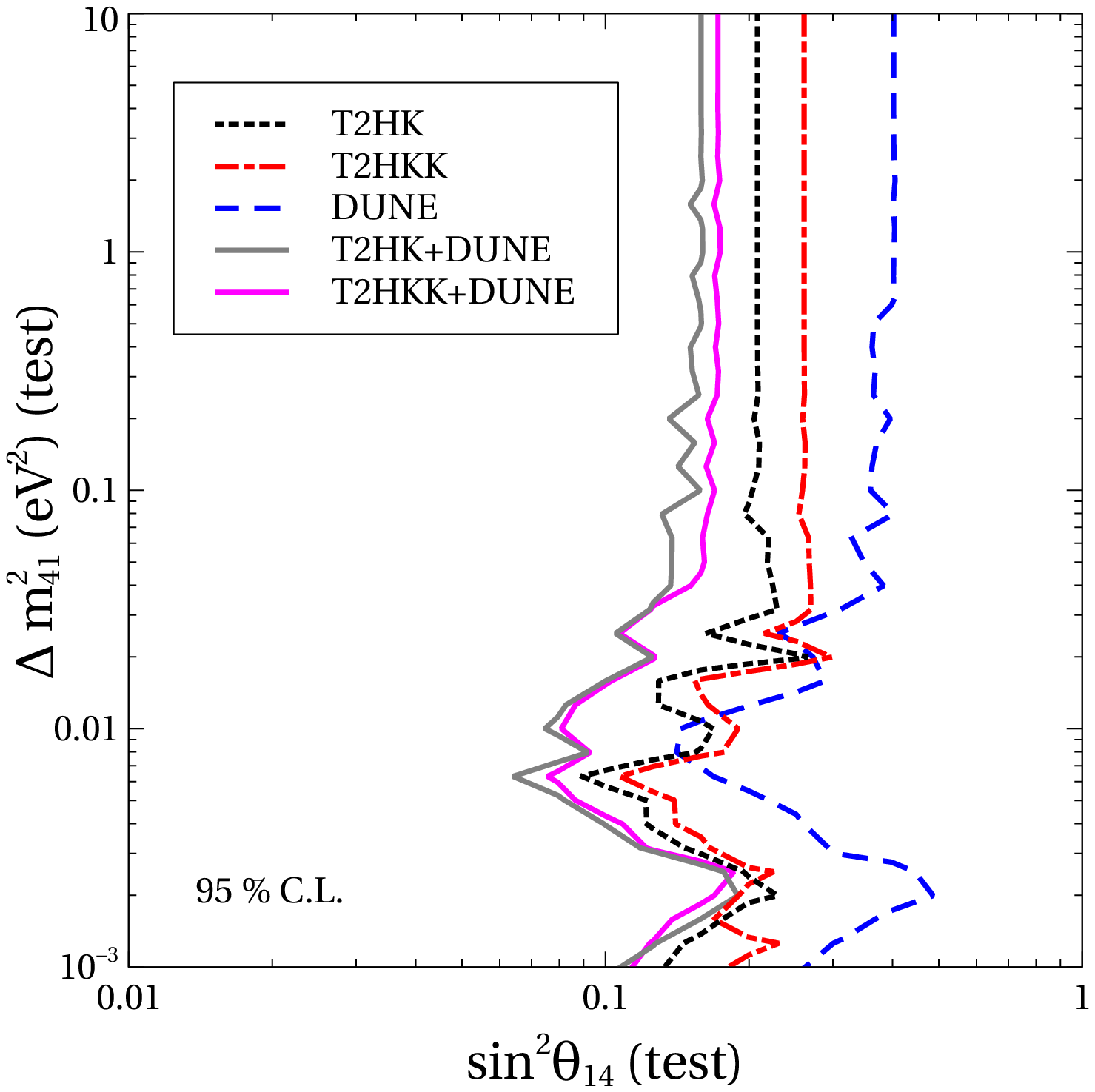}
\includegraphics[width=0.45\textwidth]{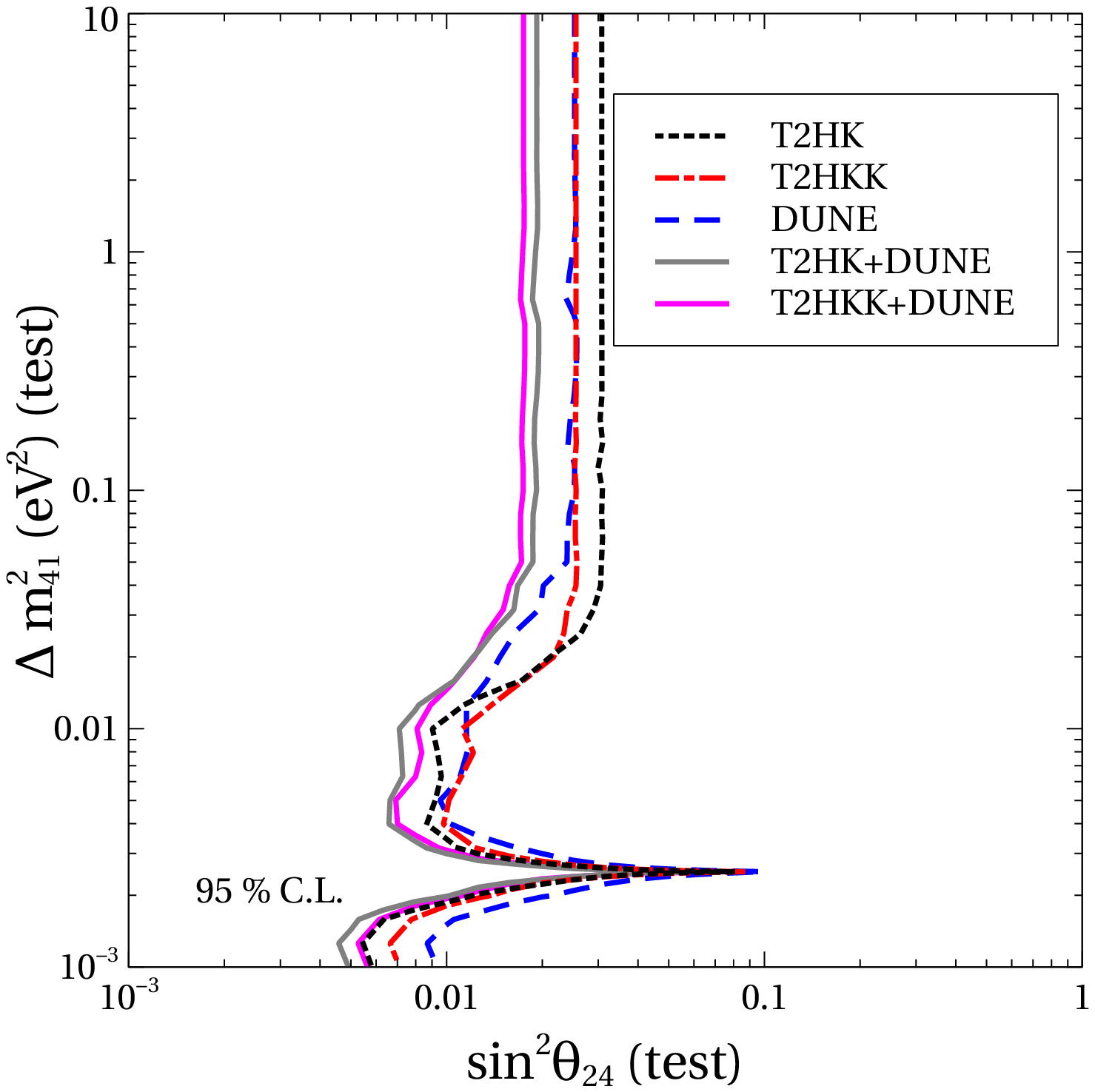}
\caption{\label{exclusion}
The expected 95 \% C.L exclusion curves in the $\sin^{2}\theta_{14}$(test)-$\Delta m^{2}_{41}$(test) plane shown in the left panel and in the $\sin^{2}\theta_{24}$(test)-$\Delta m^{2}_{41}$(test) plane shown in the right panel. The data in these plots correspond to standard three-generation oscillation scenario with no sterile mixing while the fit is done in the 3+1 framework to obtain the exclusion contours. The colour code is same as Fig.~\ref{phasen0}.}
\end{figure}

If the sterile neutrino hypothesis was wrong and there was no mixing between the active and sterile neutrinos the next-generation experiments would falsify it. There are a series of new short-baseline experiments planned which will be testing this hypothesis \citep{Choubey:2016fpi, Antonello:2015lea, Antonello:2012hf}. Even the near detector of planned long-baseline experiments are well-suited to check the sterile neutrino mixing as their baseline and energy match well to correspond to the maximum of $\Delta m_{41}^2$-driven oscillations \citep{Choubey:2016fpi}. In the same vein it is pertinent to ask how well the next-generation long-baseline experiments could constrain this hypothesis, since the oscillation probabilities for long-baseline experiments also depend on the sterile neutrino mixing and phases even though the $\Delta m_{41}^2$-driven oscillations themselves average out. While some work in this direction has already been done in the literature \citep{Berryman:2015nua, Kelly:2017kch}, we will present here, for the first time, the sensitivity of T2HKK set-up to the sterile neutrino mixing angles $\theta_{24}$ and $\theta_{14}$. We will also present the expected sensitivity from the combined prospective data-sets of T2HK (or T2HKK) and DUNE, which has not been studied before.

In Fig.~\ref{exclusion} we show the exclusion curves for the 3+1 hypothesis expected from the next-generation long-baseline experiments. The left panel of Fig.~\ref{exclusion} shows the 95 \% C.L. exclusion plots in the$\Delta m^{2}_{41}$(test)-$\sin^2\theta_{14}$(test) plane while the right panel shows the results in $\Delta m^{2}_{41}$(test)-$\sin^{2}\theta_{24}$(test)  plane. Here we generate the data assuming the standard three-generation neutrino scenario and then fit it with the 3+1 scenario. The blue dashed, black dotted and red dash-dotted curves show the exclusion plots for DUNE, T2HK and T2HKK, while the magenta and grey solid curves show the expected exclusion sensitivity for DUNE+T2HKK and DUNE+T2HK, respectively. If only three neutrinos exist in the nature, then the parameter region in the top-right of the plots are excluded at 95\% C.L. Again, as in the previous results, T2HK constrains $\theta_{14}$  better than both T2HKK and DUNE for all values of $\Delta m^2_{41}$ in the range $10^{-3}$ eV$^2$ to 10 eV$^2$. Combining DUNE with T2HK and T2HKK can improve the constraint on $\sin^2\theta_{14}$ compared to the individual experiments. Since for higher values of $\Delta m^2_{41}$ the oscillations average out, the experiments become almost insensitive to the value of $\Delta m^2_{41}$. For  $0.1$ $\rm eV^2 <\Delta m^2_{41}< 10.0$ $\rm eV^2$,  DUNE, T2HKK and T2HK can exclude $\sin^2\theta_{14} \gtap 0.4$, $\sin^2\theta_{14} \gtap 0.27$ and $\sin^2\theta_{14} \gtap 0.21$, respectively, at 95 \% C.L. For the same range of values of $\Delta m^2_{41}$, DUNE+T2HK and DUNE+T2HKK could put slightly tighter constrain on  $\sin^2\theta_{14}$ and the excluded regions are expected to be $\sin^2\theta_{14} \gtap 0.165$ and $\sin^2\theta_{14} \gtap 0.18$ at 95\% C.L., respectively.

The results presented in the right panel show the capability of these experiments to constrain $\sin^2\theta_{24}$. If $\Delta m^{2}_{41}$  is small and lies in the range  $10^{-3}$ $\rm eV^2 <\Delta m^2_{41}< 0.01$ $\rm eV^2$, T2HK gives better constraint on $\sin^2\theta_{24}$ than both DUNE and T2HKK. But for higher values of $\Delta m^{2}_{41}$, in the range of $0.014$ $\rm eV^2 <\Delta m^2_{41}< 0.1$ $\rm eV^2$, the performance of DUNE is better than both T2HK and T2HKK. For $0.1$ $\rm eV^2 <\Delta m^2_{41}< 10.0$ $\rm eV^2$, performance of T2HKK is almost similar to that of DUNE. Similar behaviour can be seen in the combined case. In the lower $\Delta m^{2}_{41}$ region, DUNE+T2HK could constrain $\sin^2\theta_{24}$ slightly better than DUNE+T2HKK. But in the higher $\Delta m^{2}_{41}$ region, DUNE+T2HKK is expected to perform better than DUNE+T2HK. The expected exclusion sensitivity for DUNE, T2HKK and T2HK in the range $0.1$ $\rm eV^2 <\Delta m^2_{41}< 10.0$ $\rm eV^2$ are given as $\sin^2\theta_{24} \gtap 0.026$,   $\sin^2\theta_{24} \gtap 0.026$ and $\sin^2\theta_{24} \gtap 0.03$,  at 95\% C.L. Similarly,  the expected exclusion limit  for DUNE+T2HKK and DUNE+T2HK at 95\% C.L. are $\sin^2\theta_{24} \gtap 0.017$ and $\sin^2\theta_{24} \gtap 0.019$, respectively, for $0.1$ $ \rm eV^2$ $<\Delta m^2_{41}$ $< 10.0$ $\rm eV^2$.

\begin{figure}
\centering
\includegraphics[width=0.5\textwidth]{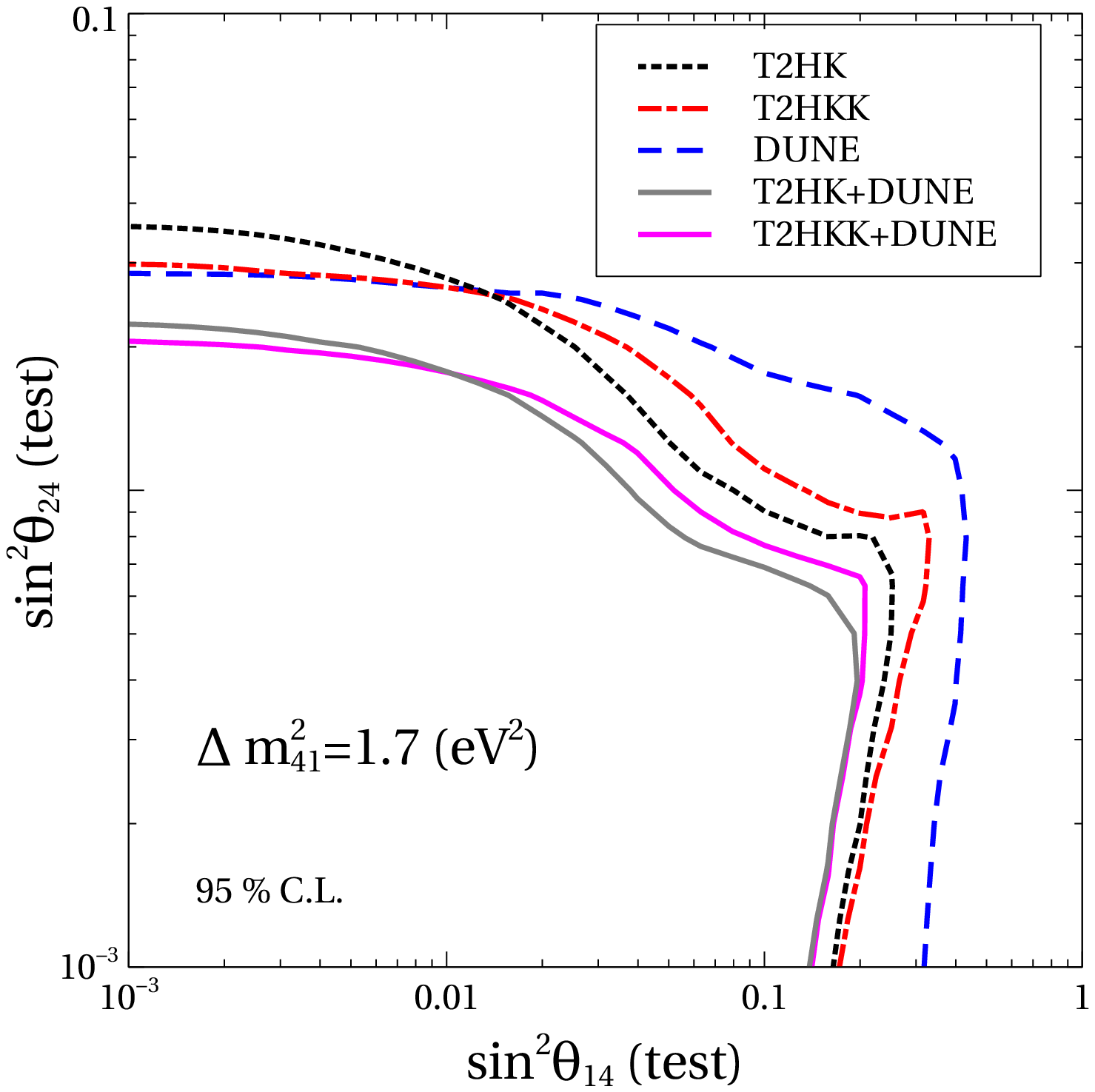}
\caption{\label{th14th24}
The expected 95 \% C.L exclusion  curves in the $\sin^{2}\theta_{14}$(test)-$\sin^{2}\theta_{24}$(test) plane for a fixed $\Delta m^{2}_{41} = 1.0$ eV$^2$. The colour code is same as Fig.~\ref{phasen0}.}
\end{figure}

In Fig.~\ref{th14th24} we show the expected exclusion contour in $\sin^{2}\theta_{14}$(test)-$\sin^{2}\theta_{24}$(test) plane. 
Here, the region outside the contour is excluded at 95\% C.L. The figure represents the slice at $\Delta m^{2}_{41}=1.7$ eV$^2$ of the contour in the $\sin^{2}_{14}$, $\sin^{2}\theta_{24}$, $\Delta m^{2}_{41}$ space. The colour code is the same as in Fig.~\ref{exclusion}. Here also, we observe better capability of T2HK to constrain $\sin^{2}\theta_{14}$-$\sin^{2}\theta_{24}$ parameter space than DUNE and T2HKK in most regions of the parameter space. The plot also shows that constraint on $\sin^2\theta_{24}$ is complicated. We see that T2HK is better than DUNE and T2HKK in constraining $\sin^2\theta_{24}$ for $\sin^2\theta_{14} \gtap 10^{-2}$. However, for $\sin^2\theta_{14} \ltap 10^{-2}$ DUNE and T2HKK perform better than T2HK in constraining $\sin^2\theta_{24} $. Combining the data-sets improves the expected sensitivity on both the sterile mixing angles.

\section{Conclusions \label{sec:conclusion}}

There are a number of observational hints that support the existence of neutrino oscillations at short baselines. Since the $\Delta m^2$ needed for these frequencies is inconsistent with the $\Delta m^2$ needed to explain the solar and atmospheric neutrino anomalies - both of which have been confirmed by earth-based experiments - it has been postulated that there are additional light neutrino states which are mixed with the three standard neutrino states. Since the $Z$-decay width restricts the number of light neutrino states coupled to the $Z$ boson to 3, the additional neutrino states should be ``sterile''. In this paper we considered one extra such sterile neutrino in the so-called 3+1 mass spectrum. In the 3+1 scenario the neutrino oscillation parameter space is extended by one new mass squared difference $\Delta m_{41}^2$, three new active-sterile mixing angles $\theta_{14}$, $\theta_{24}$ and  $\theta_{34}$ and two new CP phases $\delta_{24}$ and $\delta_{34}$. We work within a parametrisation of the mixing matrix such that the phase $\delta_{24}$ is  associated with the mixing angle $\theta_{24}$ and $\delta_{34}$ is associated with $\theta_{34}$. It is now well known that even though the $\Delta m^2_{41}$-driven oscillations are averaged out in the long-baseline experiments, the active-sterile mixing angles and the additional phases appear in the oscillation probabilities and modify it. The sensitivity of the long-baseline experiments to the active-sterile mixing angles has been studied before. The impact of the sterile neutrino parameters on the physics reach of these experiments for standard parameter measurement  such as CP violation, mass hierarchy measurement and octant of $\theta_{23}$ measurement has been investigated in details before. In this work, for the first time, we looked at the prospects of measuring the sterile CP phase $\delta_{24}$ in the long-baseline experiment T2HK (and T2HKK) and DUNE as well as when data from them is combined.

Dedicated short-baseline experiments are being built to test the active-sterile neutrino oscillation hypothesis. However, these experiments are sensitive to oscillation probabilities that have one-mass-scale-dominance. In other words, these experiments mainly work within effective two-generation scenarios and are sensitive to  $\Delta m^2_{41}$ and effective two-generation mixing angles, that can be written as combination of the mixing angles $\theta_{14}$ and $\theta_{24}$. Therefore, they are completely insensitive to the CP phases $\delta_{24}$ and $\delta_{34}$. On the other hand, these phases do affect the oscillation probabilities of the long-baseline experiments even though they are not sensitive to $\Delta m^2_{41}$ since oscillations corresponding to this frequency averages out. In particular, the probability $P_{\mu e}$ that affects the electron appearance data in long-baseline experiments depends on $\delta_{24}$. We exploited this dependence to show that the phase $\delta_{24}$ can be measured at DUNE, T2HK and T2HKK and estimated the expected precision on this parameter. Since $P_{\mu e}$ depends on $\theta_{34}$ only through earth matter effects, the effect of $\delta_{34}$ on the long-baseline data is expected to be very small and we cannot constrain it easily. Hence, in this paper we concentrated only on $\delta_{24}$. We performed a $\chi^2$ analysis of the prospective data at these experiments and presented the expected precision on $\delta_{24}$ expected at these experiments. We showed that with (7.5+2.5) years of running in (neutrino, antineutrino) mode, T2HK could constrain $\delta_{24}$ to within [$-63.0\degree$,$47.88\degree$], [$30.10\degree$, $147.9\degree$] and [$-145.95\degree,$ $-31.93\degree$] if the true value of $\delta_{24}$(true) is $0\degree$, $90\degree$ and $-90\degree$. For $\delta_{24}$(true)$=180\degree$, 
all test $\delta_{24}$ such that $\delta_{24}\leq -128.43\degree$ and $\delta_{24}\geq 136.22\degree$ are allowed in T2HK. The corresponding constraints from T2HKK and DUNE were seen to be weaker, with DUNE measurement on $\delta_{24}$ expected to be the weakest. We explained why T2HK is expected to perform better than DUNE, with the main reason being the higher statistics in T2HK. We also showed the expected allowed areas in the $\delta_{13}$(test)-$\delta_{24}$(test) plane from prospective data from DUNE, T2HK and T2HKK. We again reiterated that the expected constraints on $\delta_{24}$ were seen to be strongest from the T2HK experiment, while DUNE was seen to be weakest. We also presented the allowed areas in this plane  expected  from combined prospective data of DUNE and T2HK (or T2HKK). We showed that the combined data set could constrain the CP parameters better. Expected constraints from DUNE+T2HK was seen to be better than constraints from DUNE+T2HKK. We also discussed the impact of the sterile neutrino mixing angles and phases on the measurement of the standard CP phase $\delta_{13}$. We showed that for DUNE the expected $\delta_{13}$ precision worsens more than for T2HK. 

We also presented constraints on the mixing angles $\theta_{14}$ and $\theta_{24}$. Again, we did not consider $\theta_{34}$ since the appearance and disappearance data in long-baseline experiments depend only mildly on this angle, with the dependence coming solely from matter effects. We took two complementary approaches in this study. First we assumed the 3+1 scenario to be correct and generated data assuming non-zero values of the sterile neutrino oscillation parameters. This was used to present allowed areas in the sterile neutrino parameter space expected from full run of DUNE, T2HK, T2HKK and combinations of DUNE+T2HK and DUNE+T2HKK. We made a comparison between the expected precision reach of the different experiments. We next took the complementary approach where we assumed that the 3+1 scenario was not true. The data in this case was generated for no sterile mixing and fitted with the 3+1 hypothesis to yield expected exclusion limits on the sterile neutrino mixing parameters. Again, we did this analysis for DUNE, T2HK, T2HKK and combinations of DUNE+T2HK and DUNE+T2HKK and made a comparative analysis of the different data-sets. 

In conclusion, the sterile neutrino phases can be measured {\it only} in experiments that are sensitive to more than one oscillation frequency other than $\Delta m_{41}^2$. The long-baseline experiments therefore are the best place to measure $\delta_{24}$. We showed that the sterile phase can be measured to reasonable precision in the next-generation long-baseline experiments. The T2HK set-up is better suited to measure $\delta_{24}$ compared to DUNE due to larger statistics.

\section*{Acknowledgment}
We acknowledge the HRI cluster computing facility (http://cluster.hri.res.in). The authors would like to thank the Department of Atomic Energy (DAE) Neutrino Project of Harish-Chandra Research Institute. This project has received funding from the European Union's Horizon 2020 research and innovation programme InvisiblesPlus RISE under the Marie Sklodowska-Curie grant agreement No 690575. This project has received funding from the European Union's Horizon 2020 research and innovation programme Elusives ITN under the Marie Sklodowska- Curie grant agreement No 674896. 

\bibliography{ref}

\begin{thebibliography}{69}
\expandafter\ifx\csname natexlab\endcsname\relax\def\natexlab#1{#1}\fi
\expandafter\ifx\csname bibnamefont\endcsname\relax
  \def\bibnamefont#1{#1}\fi
\expandafter\ifx\csname bibfnamefont\endcsname\relax
  \def\bibfnamefont#1{#1}\fi
\expandafter\ifx\csname citenamefont\endcsname\relax
  \def\citenamefont#1{#1}\fi
\expandafter\ifx\csname url\endcsname\relax
  \def\url#1{\texttt{#1}}\fi
\expandafter\ifx\csname urlprefix\endcsname\relax\def\urlprefix{URL }\fi
\providecommand{\bibinfo}[2]{#2}
\providecommand{\eprint}[2][]{\url{#2}}

\bibitem[{\citenamefont{Ahmad et~al.}(2001)\citenamefont{Ahmad, Allen,
  Andersen, Anglin, B\"uhler, Barton, Beier, Bercovitch, Bigu, Biller
  et~al.}}]{PhysRevLett.87.071301}
\bibinfo{author}{\bibfnamefont{Q.~R.} \bibnamefont{Ahmad}},
  \bibinfo{author}{\bibfnamefont{R.~C.} \bibnamefont{Allen}},
  \bibinfo{author}{\bibfnamefont{T.~C.} \bibnamefont{Andersen}},
  \bibinfo{author}{\bibfnamefont{J.~D.} \bibnamefont{Anglin}},
  \bibinfo{author}{\bibfnamefont{G.}~\bibnamefont{B\"uhler}},
  \bibinfo{author}{\bibfnamefont{J.~C.} \bibnamefont{Barton}},
  \bibinfo{author}{\bibfnamefont{E.~W.} \bibnamefont{Beier}},
  \bibinfo{author}{\bibfnamefont{M.}~\bibnamefont{Bercovitch}},
  \bibinfo{author}{\bibfnamefont{J.}~\bibnamefont{Bigu}},
  \bibinfo{author}{\bibfnamefont{S.}~\bibnamefont{Biller}},
  \bibnamefont{et~al.} (\bibinfo{collaboration}{SNO Collaboration}),
  \bibinfo{journal}{Phys. Rev. Lett.} \textbf{\bibinfo{volume}{87}},
  \bibinfo{pages}{071301} (\bibinfo{year}{2001}),
  \urlprefix\url{https://link.aps.org/doi/10.1103/PhysRevLett.87.071301}.

\bibitem[{\citenamefont{Fukuda et~al.}(1998{\natexlab{a}})\citenamefont{Fukuda,
  Hayakawa, Ichihara, Inoue, Ishihara, Ishino, Itow, Kajita, Kameda, Kasuga
  et~al.}}]{PhysRevLett.81.1562}
\bibinfo{author}{\bibfnamefont{Y.}~\bibnamefont{Fukuda}},
  \bibinfo{author}{\bibfnamefont{T.}~\bibnamefont{Hayakawa}},
  \bibinfo{author}{\bibfnamefont{E.}~\bibnamefont{Ichihara}},
  \bibinfo{author}{\bibfnamefont{K.}~\bibnamefont{Inoue}},
  \bibinfo{author}{\bibfnamefont{K.}~\bibnamefont{Ishihara}},
  \bibinfo{author}{\bibfnamefont{H.}~\bibnamefont{Ishino}},
  \bibinfo{author}{\bibfnamefont{Y.}~\bibnamefont{Itow}},
  \bibinfo{author}{\bibfnamefont{T.}~\bibnamefont{Kajita}},
  \bibinfo{author}{\bibfnamefont{J.}~\bibnamefont{Kameda}},
  \bibinfo{author}{\bibfnamefont{S.}~\bibnamefont{Kasuga}},
  \bibnamefont{et~al.} (\bibinfo{collaboration}{Super-Kamiokande
  Collaboration}), \bibinfo{journal}{Phys. Rev. Lett.}
  \textbf{\bibinfo{volume}{81}}, \bibinfo{pages}{1562}
  (\bibinfo{year}{1998}{\natexlab{a}}),
  \urlprefix\url{https://link.aps.org/doi/10.1103/PhysRevLett.81.1562}.

\bibitem[{\citenamefont{Ahn et~al.}(2006)}]{Ahn:2006zza}
\bibinfo{author}{\bibfnamefont{M.~H.} \bibnamefont{Ahn}} \bibnamefont{et~al.}
  (\bibinfo{collaboration}{K2K}), \bibinfo{journal}{Phys. Rev.}
  \textbf{\bibinfo{volume}{D74}}, \bibinfo{pages}{072003}
  (\bibinfo{year}{2006}), \eprint{hep-ex/0606032}.

\bibitem[{\citenamefont{Michael et~al.}(2006)}]{Michael:2006rx}
\bibinfo{author}{\bibfnamefont{D.~G.} \bibnamefont{Michael}}
  \bibnamefont{et~al.} (\bibinfo{collaboration}{MINOS}),
  \bibinfo{journal}{Phys. Rev. Lett.} \textbf{\bibinfo{volume}{97}},
  \bibinfo{pages}{191801} (\bibinfo{year}{2006}), \eprint{hep-ex/0607088}.

\bibitem[{\citenamefont{Eguchi et~al.}(2003)}]{Eguchi:2002dm}
\bibinfo{author}{\bibfnamefont{K.}~\bibnamefont{Eguchi}} \bibnamefont{et~al.}
  (\bibinfo{collaboration}{KamLAND}), \bibinfo{journal}{Phys. Rev. Lett.}
  \textbf{\bibinfo{volume}{90}}, \bibinfo{pages}{021802}
  (\bibinfo{year}{2003}), \eprint{hep-ex/0212021}.

\bibitem[{\citenamefont{Kodama et~al.}(2001)}]{Kodama:2000mp}
\bibinfo{author}{\bibfnamefont{K.}~\bibnamefont{Kodama}} \bibnamefont{et~al.}
  (\bibinfo{collaboration}{DONUT}), \bibinfo{journal}{Phys. Lett.}
  \textbf{\bibinfo{volume}{B504}}, \bibinfo{pages}{218} (\bibinfo{year}{2001}),
  \eprint{hep-ex/0012035}.

\bibitem[{\citenamefont{Acciarri
  et~al.}(2016{\natexlab{a}})}]{Acciarri:2016crz}
\bibinfo{author}{\bibfnamefont{R.}~\bibnamefont{Acciarri}} \bibnamefont{et~al.}
  (\bibinfo{collaboration}{DUNE}) (\bibinfo{year}{2016}{\natexlab{a}}),
  \eprint{1601.05471}.

\bibitem[{\citenamefont{Acciarri et~al.}(2015)}]{Acciarri:2015uup}
\bibinfo{author}{\bibfnamefont{R.}~\bibnamefont{Acciarri}} \bibnamefont{et~al.}
  (\bibinfo{collaboration}{DUNE}) (\bibinfo{year}{2015}), \eprint{1512.06148}.

\bibitem[{\citenamefont{Strait et~al.}(2016)}]{Strait:2016mof}
\bibinfo{author}{\bibfnamefont{J.}~\bibnamefont{Strait}} \bibnamefont{et~al.}
  (\bibinfo{collaboration}{DUNE}) (\bibinfo{year}{2016}), \eprint{1601.05823}.

\bibitem[{\citenamefont{Acciarri
  et~al.}(2016{\natexlab{b}})}]{Acciarri:2016ooe}
\bibinfo{author}{\bibfnamefont{R.}~\bibnamefont{Acciarri}} \bibnamefont{et~al.}
  (\bibinfo{collaboration}{DUNE}) (\bibinfo{year}{2016}{\natexlab{b}}),
  \eprint{1601.02984}.

\bibitem[{\citenamefont{Abe et~al.}(2015)}]{Abe:2015zbg}
\bibinfo{author}{\bibfnamefont{K.}~\bibnamefont{Abe}} \bibnamefont{et~al.}
  (\bibinfo{collaboration}{Hyper-Kamiokande Proto-Collaboration}),
  \bibinfo{journal}{PTEP} \textbf{\bibinfo{volume}{2015}},
  \bibinfo{pages}{053C02} (\bibinfo{year}{2015}), \eprint{1502.05199}.

\bibitem[{\citenamefont{Abe et~al.}(2016)}]{Abe:2016ero}
\bibinfo{author}{\bibfnamefont{K.}~\bibnamefont{Abe}} \bibnamefont{et~al.}
  (\bibinfo{collaboration}{Hyper-Kamiokande proto-Collaboration})
  (\bibinfo{year}{2016}), \eprint{1611.06118}.

\bibitem[{\citenamefont{Abazajian et~al.}(2012)}]{Abazajian:2012ys}
\bibinfo{author}{\bibfnamefont{K.~N.} \bibnamefont{Abazajian}}
  \bibnamefont{et~al.} (\bibinfo{year}{2012}), \eprint{1204.5379}.

\bibitem[{\citenamefont{Athanassopoulos et~al.}(1995)}]{Athanassopoulos:1995iw}
\bibinfo{author}{\bibfnamefont{C.}~\bibnamefont{Athanassopoulos}}
  \bibnamefont{et~al.} (\bibinfo{collaboration}{LSND}), \bibinfo{journal}{Phys.
  Rev. Lett.} \textbf{\bibinfo{volume}{75}}, \bibinfo{pages}{2650}
  (\bibinfo{year}{1995}), \eprint{nucl-ex/9504002}.

\bibitem[{\citenamefont{Aguilar-Arevalo et~al.}(2001)}]{Aguilar:2001ty}
\bibinfo{author}{\bibfnamefont{A.}~\bibnamefont{Aguilar-Arevalo}}
  \bibnamefont{et~al.} (\bibinfo{collaboration}{LSND}), \bibinfo{journal}{Phys.
  Rev.} \textbf{\bibinfo{volume}{D64}}, \bibinfo{pages}{112007}
  (\bibinfo{year}{2001}), \eprint{hep-ex/0104049}.

\bibitem[{\citenamefont{Gemmeke et~al.}(1990)}]{Gemmeke:1990ix}
\bibinfo{author}{\bibfnamefont{H.}~\bibnamefont{Gemmeke}} \bibnamefont{et~al.},
  \bibinfo{journal}{Nucl. Instrum. Meth.} \textbf{\bibinfo{volume}{A289}},
  \bibinfo{pages}{490} (\bibinfo{year}{1990}).

\bibitem[{\citenamefont{Aguilar-Arevalo et~al.}(2007)}]{AguilarArevalo:2007it}
\bibinfo{author}{\bibfnamefont{A.~A.} \bibnamefont{Aguilar-Arevalo}}
  \bibnamefont{et~al.} (\bibinfo{collaboration}{MiniBooNE}),
  \bibinfo{journal}{Phys. Rev. Lett.} \textbf{\bibinfo{volume}{98}},
  \bibinfo{pages}{231801} (\bibinfo{year}{2007}), \eprint{0704.1500}.

\bibitem[{\citenamefont{Aguilar-Arevalo
  et~al.}(2013)}]{Aguilar-Arevalo:2013pmq}
\bibinfo{author}{\bibfnamefont{A.~A.} \bibnamefont{Aguilar-Arevalo}}
  \bibnamefont{et~al.} (\bibinfo{collaboration}{MiniBooNE}),
  \bibinfo{journal}{Phys. Rev. Lett.} \textbf{\bibinfo{volume}{110}},
  \bibinfo{pages}{161801} (\bibinfo{year}{2013}), \eprint{1303.2588}.

\bibitem[{\citenamefont{Aguilar-Arevalo et~al.}(2010)}]{AguilarArevalo:2010wv}
\bibinfo{author}{\bibfnamefont{A.~A.} \bibnamefont{Aguilar-Arevalo}}
  \bibnamefont{et~al.} (\bibinfo{collaboration}{MiniBooNE}),
  \bibinfo{journal}{Phys. Rev. Lett.} \textbf{\bibinfo{volume}{105}},
  \bibinfo{pages}{181801} (\bibinfo{year}{2010}), \eprint{1007.1150}.

\bibitem[{\citenamefont{Goswami}(1997)}]{Goswami:1995yq}
\bibinfo{author}{\bibfnamefont{S.}~\bibnamefont{Goswami}},
  \bibinfo{journal}{Phys. Rev.} \textbf{\bibinfo{volume}{D55}},
  \bibinfo{pages}{2931} (\bibinfo{year}{1997}), \eprint{hep-ph/9507212}.

\bibitem[{\citenamefont{Gariazzo et~al.}(2017)\citenamefont{Gariazzo, Giunti,
  Laveder, and Li}}]{Gariazzo:2017fdh}
\bibinfo{author}{\bibfnamefont{S.}~\bibnamefont{Gariazzo}},
  \bibinfo{author}{\bibfnamefont{C.}~\bibnamefont{Giunti}},
  \bibinfo{author}{\bibfnamefont{M.}~\bibnamefont{Laveder}}, \bibnamefont{and}
  \bibinfo{author}{\bibfnamefont{Y.~F.} \bibnamefont{Li}},
  \bibinfo{journal}{JHEP} \textbf{\bibinfo{volume}{06}}, \bibinfo{pages}{135}
  (\bibinfo{year}{2017}), \eprint{1703.00860}.

\bibitem[{\citenamefont{Gandhi et~al.}(2015)\citenamefont{Gandhi, Kayser,
  Masud, and Prakash}}]{Gandhi:2015xza}
\bibinfo{author}{\bibfnamefont{R.}~\bibnamefont{Gandhi}},
  \bibinfo{author}{\bibfnamefont{B.}~\bibnamefont{Kayser}},
  \bibinfo{author}{\bibfnamefont{M.}~\bibnamefont{Masud}}, \bibnamefont{and}
  \bibinfo{author}{\bibfnamefont{S.}~\bibnamefont{Prakash}},
  \bibinfo{journal}{JHEP} \textbf{\bibinfo{volume}{11}}, \bibinfo{pages}{039}
  (\bibinfo{year}{2015}), \eprint{1508.06275}.

\bibitem[{\citenamefont{Dutta et~al.}(2016)\citenamefont{Dutta, Gandhi, Kayser,
  Masud, and Prakash}}]{Dutta:2016glq}
\bibinfo{author}{\bibfnamefont{D.}~\bibnamefont{Dutta}},
  \bibinfo{author}{\bibfnamefont{R.}~\bibnamefont{Gandhi}},
  \bibinfo{author}{\bibfnamefont{B.}~\bibnamefont{Kayser}},
  \bibinfo{author}{\bibfnamefont{M.}~\bibnamefont{Masud}}, \bibnamefont{and}
  \bibinfo{author}{\bibfnamefont{S.}~\bibnamefont{Prakash}},
  \bibinfo{journal}{JHEP} \textbf{\bibinfo{volume}{11}}, \bibinfo{pages}{122}
  (\bibinfo{year}{2016}), \eprint{1607.02152}.

\bibitem[{\citenamefont{Berryman et~al.}(2015)\citenamefont{Berryman,
  de~Gouvêa, Kelly, and Kobach}}]{Berryman:2015nua}
\bibinfo{author}{\bibfnamefont{J.~M.} \bibnamefont{Berryman}},
  \bibinfo{author}{\bibfnamefont{A.}~\bibnamefont{de~Gouvêa}},
  \bibinfo{author}{\bibfnamefont{K.~J.} \bibnamefont{Kelly}}, \bibnamefont{and}
  \bibinfo{author}{\bibfnamefont{A.}~\bibnamefont{Kobach}},
  \bibinfo{journal}{Phys. Rev.} \textbf{\bibinfo{volume}{D92}},
  \bibinfo{pages}{073012} (\bibinfo{year}{2015}), \eprint{1507.03986}.

\bibitem[{\citenamefont{de~Gouvêa et~al.}(2015)\citenamefont{de~Gouvêa, Kelly,
  and Kobach}}]{deGouvea:2014aoa}
\bibinfo{author}{\bibfnamefont{A.}~\bibnamefont{de~Gouvêa}},
  \bibinfo{author}{\bibfnamefont{K.~J.} \bibnamefont{Kelly}}, \bibnamefont{and}
  \bibinfo{author}{\bibfnamefont{A.}~\bibnamefont{Kobach}},
  \bibinfo{journal}{Phys. Rev.} \textbf{\bibinfo{volume}{D91}},
  \bibinfo{pages}{053005} (\bibinfo{year}{2015}), \eprint{1412.1479}.

\bibitem[{\citenamefont{Agarwalla et~al.}(2017)\citenamefont{Agarwalla,
  Chatterjee, and Palazzo}}]{Agarwalla:2016xlg}
\bibinfo{author}{\bibfnamefont{S.~K.} \bibnamefont{Agarwalla}},
  \bibinfo{author}{\bibfnamefont{S.~S.} \bibnamefont{Chatterjee}},
  \bibnamefont{and} \bibinfo{author}{\bibfnamefont{A.}~\bibnamefont{Palazzo}},
  \bibinfo{journal}{Phys. Rev. Lett.} \textbf{\bibinfo{volume}{118}},
  \bibinfo{pages}{031804} (\bibinfo{year}{2017}), \eprint{1605.04299}.

\bibitem[{\citenamefont{Agarwalla
  et~al.}(2016{\natexlab{a}})\citenamefont{Agarwalla, Chatterjee, and
  Palazzo}}]{Agarwalla:2016xxa}
\bibinfo{author}{\bibfnamefont{S.~K.} \bibnamefont{Agarwalla}},
  \bibinfo{author}{\bibfnamefont{S.~S.} \bibnamefont{Chatterjee}},
  \bibnamefont{and} \bibinfo{author}{\bibfnamefont{A.}~\bibnamefont{Palazzo}},
  \bibinfo{journal}{JHEP} \textbf{\bibinfo{volume}{09}}, \bibinfo{pages}{016}
  (\bibinfo{year}{2016}{\natexlab{a}}), \eprint{1603.03759}.

\bibitem[{\citenamefont{Agarwalla
  et~al.}(2016{\natexlab{b}})\citenamefont{Agarwalla, Chatterjee, Dasgupta, and
  Palazzo}}]{Agarwalla:2016mrc}
\bibinfo{author}{\bibfnamefont{S.~K.} \bibnamefont{Agarwalla}},
  \bibinfo{author}{\bibfnamefont{S.~S.} \bibnamefont{Chatterjee}},
  \bibinfo{author}{\bibfnamefont{A.}~\bibnamefont{Dasgupta}}, \bibnamefont{and}
  \bibinfo{author}{\bibfnamefont{A.}~\bibnamefont{Palazzo}},
  \bibinfo{journal}{JHEP} \textbf{\bibinfo{volume}{02}}, \bibinfo{pages}{111}
  (\bibinfo{year}{2016}{\natexlab{b}}), \eprint{1601.05995}.

\bibitem[{\citenamefont{Klop and Palazzo}(2015)}]{Klop:2014ima}
\bibinfo{author}{\bibfnamefont{N.}~\bibnamefont{Klop}} \bibnamefont{and}
  \bibinfo{author}{\bibfnamefont{A.}~\bibnamefont{Palazzo}},
  \bibinfo{journal}{Phys. Rev.} \textbf{\bibinfo{volume}{D91}},
  \bibinfo{pages}{073017} (\bibinfo{year}{2015}), \eprint{1412.7524}.

\bibitem[{\citenamefont{Choubey
  et~al.}(2017{\natexlab{a}})\citenamefont{Choubey, Dutta, and
  Pramanik}}]{Choubey:2017cba}
\bibinfo{author}{\bibfnamefont{S.}~\bibnamefont{Choubey}},
  \bibinfo{author}{\bibfnamefont{D.}~\bibnamefont{Dutta}}, \bibnamefont{and}
  \bibinfo{author}{\bibfnamefont{D.}~\bibnamefont{Pramanik}},
  \bibinfo{journal}{Phys. Rev.} \textbf{\bibinfo{volume}{D96}},
  \bibinfo{pages}{056026} (\bibinfo{year}{2017}{\natexlab{a}}),
  \eprint{1704.07269}.

\bibitem[{\citenamefont{Ghosh et~al.}(2017)\citenamefont{Ghosh, Gupta,
  Matthews, Sharma, and Williams}}]{Ghosh:2017atj}
\bibinfo{author}{\bibfnamefont{M.}~\bibnamefont{Ghosh}},
  \bibinfo{author}{\bibfnamefont{S.}~\bibnamefont{Gupta}},
  \bibinfo{author}{\bibfnamefont{Z.~M.} \bibnamefont{Matthews}},
  \bibinfo{author}{\bibfnamefont{P.}~\bibnamefont{Sharma}}, \bibnamefont{and}
  \bibinfo{author}{\bibfnamefont{A.~G.} \bibnamefont{Williams}},
  \bibinfo{journal}{Phys. Rev.} \textbf{\bibinfo{volume}{D96}},
  \bibinfo{pages}{075018} (\bibinfo{year}{2017}), \eprint{1704.04771}.

\bibitem[{\citenamefont{Masud et~al.}(2016)\citenamefont{Masud, Chatterjee, and
  Mehta}}]{Masud:2015xva}
\bibinfo{author}{\bibfnamefont{M.}~\bibnamefont{Masud}},
  \bibinfo{author}{\bibfnamefont{A.}~\bibnamefont{Chatterjee}},
  \bibnamefont{and} \bibinfo{author}{\bibfnamefont{P.}~\bibnamefont{Mehta}},
  \bibinfo{journal}{J. Phys.} \textbf{\bibinfo{volume}{G43}},
  \bibinfo{pages}{095005} (\bibinfo{year}{2016}), \eprint{1510.08261}.

\bibitem[{\citenamefont{de~Gouvêa and Kelly}(2016)}]{deGouvea:2015ndi}
\bibinfo{author}{\bibfnamefont{A.}~\bibnamefont{de~Gouvêa}} \bibnamefont{and}
  \bibinfo{author}{\bibfnamefont{K.~J.} \bibnamefont{Kelly}},
  \bibinfo{journal}{Nucl. Phys.} \textbf{\bibinfo{volume}{B908}},
  \bibinfo{pages}{318} (\bibinfo{year}{2016}), \eprint{1511.05562}.

\bibitem[{\citenamefont{Coloma}(2016)}]{Coloma:2015kiu}
\bibinfo{author}{\bibfnamefont{P.}~\bibnamefont{Coloma}},
  \bibinfo{journal}{JHEP} \textbf{\bibinfo{volume}{03}}, \bibinfo{pages}{016}
  (\bibinfo{year}{2016}), \eprint{1511.06357}.

\bibitem[{\citenamefont{Liao et~al.}(2016)\citenamefont{Liao, Marfatia, and
  Whisnant}}]{Liao:2016hsa}
\bibinfo{author}{\bibfnamefont{J.}~\bibnamefont{Liao}},
  \bibinfo{author}{\bibfnamefont{D.}~\bibnamefont{Marfatia}}, \bibnamefont{and}
  \bibinfo{author}{\bibfnamefont{K.}~\bibnamefont{Whisnant}},
  \bibinfo{journal}{Phys. Rev.} \textbf{\bibinfo{volume}{D93}},
  \bibinfo{pages}{093016} (\bibinfo{year}{2016}), \eprint{1601.00927}.

\bibitem[{\citenamefont{Masud and Mehta}(2016)}]{Masud:2016bvp}
\bibinfo{author}{\bibfnamefont{M.}~\bibnamefont{Masud}} \bibnamefont{and}
  \bibinfo{author}{\bibfnamefont{P.}~\bibnamefont{Mehta}},
  \bibinfo{journal}{Phys. Rev.} \textbf{\bibinfo{volume}{D94}},
  \bibinfo{pages}{013014} (\bibinfo{year}{2016}), \eprint{1603.01380}.

\bibitem[{\citenamefont{Agarwalla
  et~al.}(2016{\natexlab{c}})\citenamefont{Agarwalla, Chatterjee, and
  Palazzo}}]{Agarwalla:2016fkh}
\bibinfo{author}{\bibfnamefont{S.~K.} \bibnamefont{Agarwalla}},
  \bibinfo{author}{\bibfnamefont{S.~S.} \bibnamefont{Chatterjee}},
  \bibnamefont{and} \bibinfo{author}{\bibfnamefont{A.}~\bibnamefont{Palazzo}},
  \bibinfo{journal}{Phys. Lett.} \textbf{\bibinfo{volume}{B762}},
  \bibinfo{pages}{64} (\bibinfo{year}{2016}{\natexlab{c}}),
  \eprint{1607.01745}.

\bibitem[{\citenamefont{Blennow et~al.}(2016)\citenamefont{Blennow, Choubey,
  Ohlsson, Pramanik, and Raut}}]{Blennow:2016etl}
\bibinfo{author}{\bibfnamefont{M.}~\bibnamefont{Blennow}},
  \bibinfo{author}{\bibfnamefont{S.}~\bibnamefont{Choubey}},
  \bibinfo{author}{\bibfnamefont{T.}~\bibnamefont{Ohlsson}},
  \bibinfo{author}{\bibfnamefont{D.}~\bibnamefont{Pramanik}}, \bibnamefont{and}
  \bibinfo{author}{\bibfnamefont{S.~K.} \bibnamefont{Raut}},
  \bibinfo{journal}{JHEP} \textbf{\bibinfo{volume}{08}}, \bibinfo{pages}{090}
  (\bibinfo{year}{2016}), \eprint{1606.08851}.

\bibitem[{\citenamefont{Deepthi et~al.}(2017)\citenamefont{Deepthi, Goswami,
  and Nath}}]{Deepthi:2017gxg}
\bibinfo{author}{\bibfnamefont{K.~N.} \bibnamefont{Deepthi}},
  \bibinfo{author}{\bibfnamefont{S.}~\bibnamefont{Goswami}}, \bibnamefont{and}
  \bibinfo{author}{\bibfnamefont{N.}~\bibnamefont{Nath}}
  (\bibinfo{year}{2017}), \eprint{1711.04840}.

\bibitem[{\citenamefont{Liao et~al.}(2017)\citenamefont{Liao, Marfatia, and
  Whisnant}}]{Liao:2016orc}
\bibinfo{author}{\bibfnamefont{J.}~\bibnamefont{Liao}},
  \bibinfo{author}{\bibfnamefont{D.}~\bibnamefont{Marfatia}}, \bibnamefont{and}
  \bibinfo{author}{\bibfnamefont{K.}~\bibnamefont{Whisnant}},
  \bibinfo{journal}{JHEP} \textbf{\bibinfo{volume}{01}}, \bibinfo{pages}{071}
  (\bibinfo{year}{2017}), \eprint{1612.01443}.

\bibitem[{\citenamefont{Fukasawa et~al.}(2017)\citenamefont{Fukasawa, Ghosh,
  and Yasuda}}]{Fukasawa:2016lew}
\bibinfo{author}{\bibfnamefont{S.}~\bibnamefont{Fukasawa}},
  \bibinfo{author}{\bibfnamefont{M.}~\bibnamefont{Ghosh}}, \bibnamefont{and}
  \bibinfo{author}{\bibfnamefont{O.}~\bibnamefont{Yasuda}},
  \bibinfo{journal}{Phys. Rev.} \textbf{\bibinfo{volume}{D95}},
  \bibinfo{pages}{055005} (\bibinfo{year}{2017}), \eprint{1611.06141}.

\bibitem[{\citenamefont{Dutta and Ghoshal}(2016)}]{Dutta:2016vcc}
\bibinfo{author}{\bibfnamefont{D.}~\bibnamefont{Dutta}} \bibnamefont{and}
  \bibinfo{author}{\bibfnamefont{P.}~\bibnamefont{Ghoshal}},
  \bibinfo{journal}{JHEP} \textbf{\bibinfo{volume}{09}}, \bibinfo{pages}{110}
  (\bibinfo{year}{2016}), \eprint{1607.02500}.

\bibitem[{\citenamefont{Dutta et~al.}(2017{\natexlab{a}})\citenamefont{Dutta,
  Ghoshal, and Roy}}]{Dutta:2016czj}
\bibinfo{author}{\bibfnamefont{D.}~\bibnamefont{Dutta}},
  \bibinfo{author}{\bibfnamefont{P.}~\bibnamefont{Ghoshal}}, \bibnamefont{and}
  \bibinfo{author}{\bibfnamefont{S.}~\bibnamefont{Roy}},
  \bibinfo{journal}{Nucl. Phys.} \textbf{\bibinfo{volume}{B920}},
  \bibinfo{pages}{385} (\bibinfo{year}{2017}{\natexlab{a}}),
  \eprint{1609.07094}.

\bibitem[{\citenamefont{Blennow et~al.}(2017)\citenamefont{Blennow, Coloma,
  Fernandez-Martinez, Hernandez-Garcia, and Lopez-Pavon}}]{Blennow:2016jkn}
\bibinfo{author}{\bibfnamefont{M.}~\bibnamefont{Blennow}},
  \bibinfo{author}{\bibfnamefont{P.}~\bibnamefont{Coloma}},
  \bibinfo{author}{\bibfnamefont{E.}~\bibnamefont{Fernandez-Martinez}},
  \bibinfo{author}{\bibfnamefont{J.}~\bibnamefont{Hernandez-Garcia}},
  \bibnamefont{and}
  \bibinfo{author}{\bibfnamefont{J.}~\bibnamefont{Lopez-Pavon}},
  \bibinfo{journal}{JHEP} \textbf{\bibinfo{volume}{04}}, \bibinfo{pages}{153}
  (\bibinfo{year}{2017}), \eprint{1609.08637}.

\bibitem[{\citenamefont{Dutta et~al.}(2017{\natexlab{b}})\citenamefont{Dutta,
  Ghoshal, and Sehrawat}}]{Dutta:2016eks}
\bibinfo{author}{\bibfnamefont{D.}~\bibnamefont{Dutta}},
  \bibinfo{author}{\bibfnamefont{P.}~\bibnamefont{Ghoshal}}, \bibnamefont{and}
  \bibinfo{author}{\bibfnamefont{S.~K.} \bibnamefont{Sehrawat}},
  \bibinfo{journal}{Phys. Rev.} \textbf{\bibinfo{volume}{D95}},
  \bibinfo{pages}{095007} (\bibinfo{year}{2017}{\natexlab{b}}),
  \eprint{1610.07203}.

\bibitem[{\citenamefont{Escrihuela et~al.}(2017)\citenamefont{Escrihuela,
  Forero, Miranda, Tórtola, and Valle}}]{Escrihuela:2016ube}
\bibinfo{author}{\bibfnamefont{F.~J.} \bibnamefont{Escrihuela}},
  \bibinfo{author}{\bibfnamefont{D.~V.} \bibnamefont{Forero}},
  \bibinfo{author}{\bibfnamefont{O.~G.} \bibnamefont{Miranda}},
  \bibinfo{author}{\bibfnamefont{M.}~\bibnamefont{Tórtola}}, \bibnamefont{and}
  \bibinfo{author}{\bibfnamefont{J.~W.~F.} \bibnamefont{Valle}},
  \bibinfo{journal}{New J. Phys.} \textbf{\bibinfo{volume}{19}},
  \bibinfo{pages}{093005} (\bibinfo{year}{2017}), \eprint{1612.07377}.

\bibitem[{\citenamefont{Ge et~al.}(2017)\citenamefont{Ge, Pasquini, Tortola,
  and Valle}}]{Ge:2016xya}
\bibinfo{author}{\bibfnamefont{S.-F.} \bibnamefont{Ge}},
  \bibinfo{author}{\bibfnamefont{P.}~\bibnamefont{Pasquini}},
  \bibinfo{author}{\bibfnamefont{M.}~\bibnamefont{Tortola}}, \bibnamefont{and}
  \bibinfo{author}{\bibfnamefont{J.~W.~F.} \bibnamefont{Valle}},
  \bibinfo{journal}{Phys. Rev.} \textbf{\bibinfo{volume}{D95}},
  \bibinfo{pages}{033005} (\bibinfo{year}{2017}), \eprint{1605.01670}.

\bibitem[{\citenamefont{Escrihuela et~al.}(2015)\citenamefont{Escrihuela,
  Forero, Miranda, Tortola, and Valle}}]{Escrihuela:2015wra}
\bibinfo{author}{\bibfnamefont{F.~J.} \bibnamefont{Escrihuela}},
  \bibinfo{author}{\bibfnamefont{D.~V.} \bibnamefont{Forero}},
  \bibinfo{author}{\bibfnamefont{O.~G.} \bibnamefont{Miranda}},
  \bibinfo{author}{\bibfnamefont{M.}~\bibnamefont{Tortola}}, \bibnamefont{and}
  \bibinfo{author}{\bibfnamefont{J.~W.~F.} \bibnamefont{Valle}},
  \bibinfo{journal}{Phys. Rev.} \textbf{\bibinfo{volume}{D92}},
  \bibinfo{pages}{053009} (\bibinfo{year}{2015}), \bibinfo{note}{[Erratum:
  Phys. Rev.D93,no.11,119905(2016)]}, \eprint{1503.08879}.

\bibitem[{\citenamefont{Ge and Smirnov}(2016)}]{Ge:2016dlx}
\bibinfo{author}{\bibfnamefont{S.-F.} \bibnamefont{Ge}} \bibnamefont{and}
  \bibinfo{author}{\bibfnamefont{A.~{\relax Yu}.} \bibnamefont{Smirnov}},
  \bibinfo{journal}{JHEP} \textbf{\bibinfo{volume}{10}}, \bibinfo{pages}{138}
  (\bibinfo{year}{2016}), \eprint{1607.08513}.

\bibitem[{\citenamefont{Tang et~al.}(2017)\citenamefont{Tang, Zhang, and
  Li}}]{Tang:2017khg}
\bibinfo{author}{\bibfnamefont{J.}~\bibnamefont{Tang}},
  \bibinfo{author}{\bibfnamefont{Y.}~\bibnamefont{Zhang}}, \bibnamefont{and}
  \bibinfo{author}{\bibfnamefont{Y.-F.} \bibnamefont{Li}},
  \bibinfo{journal}{Phys. Lett.} \textbf{\bibinfo{volume}{B774}},
  \bibinfo{pages}{217} (\bibinfo{year}{2017}), \eprint{1708.04909}.

\bibitem[{\citenamefont{Choubey
  et~al.}(2017{\natexlab{b}})\citenamefont{Choubey, Goswami, and
  Pramanik}}]{Choubey:2017dyu}
\bibinfo{author}{\bibfnamefont{S.}~\bibnamefont{Choubey}},
  \bibinfo{author}{\bibfnamefont{S.}~\bibnamefont{Goswami}}, \bibnamefont{and}
  \bibinfo{author}{\bibfnamefont{D.}~\bibnamefont{Pramanik}}
  (\bibinfo{year}{2017}{\natexlab{b}}), \eprint{1705.05820}.

\bibitem[{\citenamefont{Chatterjee et~al.}(2015)\citenamefont{Chatterjee,
  Dasgupta, and Agarwalla}}]{Chatterjee:2015gta}
\bibinfo{author}{\bibfnamefont{S.~S.} \bibnamefont{Chatterjee}},
  \bibinfo{author}{\bibfnamefont{A.}~\bibnamefont{Dasgupta}}, \bibnamefont{and}
  \bibinfo{author}{\bibfnamefont{S.~K.} \bibnamefont{Agarwalla}},
  \bibinfo{journal}{JHEP} \textbf{\bibinfo{volume}{12}}, \bibinfo{pages}{167}
  (\bibinfo{year}{2015}), \eprint{1509.03517}.

\bibitem[{\citenamefont{Berryman et~al.}(2016)\citenamefont{Berryman,
  de~Gouvêa, Kelly, Peres, and Tabrizi}}]{Berryman:2016szd}
\bibinfo{author}{\bibfnamefont{J.~M.} \bibnamefont{Berryman}},
  \bibinfo{author}{\bibfnamefont{A.}~\bibnamefont{de~Gouvêa}},
  \bibinfo{author}{\bibfnamefont{K.~J.} \bibnamefont{Kelly}},
  \bibinfo{author}{\bibfnamefont{O.~L.~G.} \bibnamefont{Peres}},
  \bibnamefont{and} \bibinfo{author}{\bibfnamefont{Z.}~\bibnamefont{Tabrizi}},
  \bibinfo{journal}{Phys. Rev.} \textbf{\bibinfo{volume}{D94}},
  \bibinfo{pages}{033006} (\bibinfo{year}{2016}), \eprint{1603.00018}.

\bibitem[{\citenamefont{Antonello et~al.}(2015)}]{Antonello:2015lea}
\bibinfo{author}{\bibfnamefont{M.}~\bibnamefont{Antonello}}
  \bibnamefont{et~al.} (\bibinfo{collaboration}{LAr1-ND, ICARUS-WA104,
  MicroBooNE}) (\bibinfo{year}{2015}), \eprint{1503.01520}.

\bibitem[{\citenamefont{Antonello et~al.}(2012)}]{Antonello:2012hf}
\bibinfo{author}{\bibfnamefont{M.}~\bibnamefont{Antonello}}
  \bibnamefont{et~al.} (\bibinfo{year}{2012}), \eprint{1203.3432}.

\bibitem[{\citenamefont{Jones}(2013)}]{Jones:2011ci}
\bibinfo{author}{\bibfnamefont{B.~J.~P.} \bibnamefont{Jones}},
  \bibinfo{journal}{J. Phys. Conf. Ser.} \textbf{\bibinfo{volume}{408}},
  \bibinfo{pages}{012028} (\bibinfo{year}{2013}), \eprint{1110.1678}.

\bibitem[{\citenamefont{Choubey and Pramanik}(2017)}]{Choubey:2016fpi}
\bibinfo{author}{\bibfnamefont{S.}~\bibnamefont{Choubey}} \bibnamefont{and}
  \bibinfo{author}{\bibfnamefont{D.}~\bibnamefont{Pramanik}},
  \bibinfo{journal}{Phys. Lett.} \textbf{\bibinfo{volume}{B764}},
  \bibinfo{pages}{135} (\bibinfo{year}{2017}), \eprint{1604.04731}.

\bibitem[{\citenamefont{Kelly}(2017)}]{Kelly:2017kch}
\bibinfo{author}{\bibfnamefont{K.~J.} \bibnamefont{Kelly}},
  \bibinfo{journal}{Phys. Rev.} \textbf{\bibinfo{volume}{D95}},
  \bibinfo{pages}{115009} (\bibinfo{year}{2017}), \eprint{1703.00448}.

\bibitem[{\citenamefont{Huber et~al.}(2005)\citenamefont{Huber, Lindner, and
  Winter}}]{Huber:2004ka}
\bibinfo{author}{\bibfnamefont{P.}~\bibnamefont{Huber}},
  \bibinfo{author}{\bibfnamefont{M.}~\bibnamefont{Lindner}}, \bibnamefont{and}
  \bibinfo{author}{\bibfnamefont{W.}~\bibnamefont{Winter}},
  \bibinfo{journal}{Comput. Phys. Commun.} \textbf{\bibinfo{volume}{167}},
  \bibinfo{pages}{195} (\bibinfo{year}{2005}), \eprint{hep-ph/0407333}.

\bibitem[{\citenamefont{Huber et~al.}(2007)\citenamefont{Huber, Kopp, Lindner,
  Rolinec, and Winter}}]{Huber:2007ji}
\bibinfo{author}{\bibfnamefont{P.}~\bibnamefont{Huber}},
  \bibinfo{author}{\bibfnamefont{J.}~\bibnamefont{Kopp}},
  \bibinfo{author}{\bibfnamefont{M.}~\bibnamefont{Lindner}},
  \bibinfo{author}{\bibfnamefont{M.}~\bibnamefont{Rolinec}}, \bibnamefont{and}
  \bibinfo{author}{\bibfnamefont{W.}~\bibnamefont{Winter}},
  \bibinfo{journal}{Comput. Phys. Commun.} \textbf{\bibinfo{volume}{177}},
  \bibinfo{pages}{432} (\bibinfo{year}{2007}), \eprint{hep-ph/0701187}.

\bibitem[{\citenamefont{Kopp}(2008)}]{Kopp:2006wp}
\bibinfo{author}{\bibfnamefont{J.}~\bibnamefont{Kopp}}, \bibinfo{journal}{Int.
  J. Mod. Phys.} \textbf{\bibinfo{volume}{C19}}, \bibinfo{pages}{523}
  (\bibinfo{year}{2008}), \bibinfo{note}{erratum ibid.\ {\bf C19} (2008) 845},
  \eprint{physics/0610206}.

\bibitem[{\citenamefont{Kopp et~al.}(2008)\citenamefont{Kopp, Lindner, Ota, and
  Sato}}]{Kopp:2007ne}
\bibinfo{author}{\bibfnamefont{J.}~\bibnamefont{Kopp}},
  \bibinfo{author}{\bibfnamefont{M.}~\bibnamefont{Lindner}},
  \bibinfo{author}{\bibfnamefont{T.}~\bibnamefont{Ota}}, \bibnamefont{and}
  \bibinfo{author}{\bibfnamefont{J.}~\bibnamefont{Sato}},
  \bibinfo{journal}{Phys. Rev.} \textbf{\bibinfo{volume}{D77}},
  \bibinfo{pages}{013007} (\bibinfo{year}{2008}), \eprint{0708.0152}.

\bibitem[{\citenamefont{Esteban et~al.}(2017)\citenamefont{Esteban,
  Gonzalez-Garcia, Maltoni, Martinez-Soler, and Schwetz}}]{Esteban:2016qun}
\bibinfo{author}{\bibfnamefont{I.}~\bibnamefont{Esteban}},
  \bibinfo{author}{\bibfnamefont{M.~C.} \bibnamefont{Gonzalez-Garcia}},
  \bibinfo{author}{\bibfnamefont{M.}~\bibnamefont{Maltoni}},
  \bibinfo{author}{\bibfnamefont{I.}~\bibnamefont{Martinez-Soler}},
  \bibnamefont{and} \bibinfo{author}{\bibfnamefont{T.}~\bibnamefont{Schwetz}},
  \bibinfo{journal}{JHEP} \textbf{\bibinfo{volume}{01}}, \bibinfo{pages}{087}
  (\bibinfo{year}{2017}), \eprint{1611.01514}.

\bibitem[{\citenamefont{Alion et~al.}(2016)}]{Alion:2016uaj}
\bibinfo{author}{\bibfnamefont{T.}~\bibnamefont{Alion}} \bibnamefont{et~al.}
  (\bibinfo{collaboration}{DUNE}) (\bibinfo{year}{2016}), \eprint{1606.09550}.

\bibitem[{\citenamefont{Fukuda et~al.}(1998{\natexlab{b}})}]{Fukuda:1998mi}
\bibinfo{author}{\bibfnamefont{Y.}~\bibnamefont{Fukuda}} \bibnamefont{et~al.}
  (\bibinfo{collaboration}{Super-Kamiokande}), \bibinfo{journal}{Phys. Rev.
  Lett.} \textbf{\bibinfo{volume}{81}}, \bibinfo{pages}{1562}
  (\bibinfo{year}{1998}{\natexlab{b}}), \eprint{hep-ex/9807003}.

\bibitem[{\citenamefont{Abe et~al.}(2011)}]{Abe:2011ks}
\bibinfo{author}{\bibfnamefont{K.}~\bibnamefont{Abe}} \bibnamefont{et~al.}
  (\bibinfo{collaboration}{T2K}), \bibinfo{journal}{Nucl. Instrum. Meth.}
  \textbf{\bibinfo{volume}{A659}}, \bibinfo{pages}{106} (\bibinfo{year}{2011}),
  \eprint{1106.1238}.

\bibitem[{\citenamefont{Adamson et~al.}(2017)}]{Adamson:2017zcg}
\bibinfo{author}{\bibfnamefont{P.}~\bibnamefont{Adamson}} \bibnamefont{et~al.}
  (\bibinfo{collaboration}{NOvA}), \bibinfo{journal}{Phys. Rev.}
  \textbf{\bibinfo{volume}{D96}}, \bibinfo{pages}{072006}
  (\bibinfo{year}{2017}), \eprint{1706.04592}.

\bibitem[{\citenamefont{Gandhi et~al.}(2017)\citenamefont{Gandhi, Kayser,
  Prakash, and Roy}}]{Gandhi:2017vzo}
\bibinfo{author}{\bibfnamefont{R.}~\bibnamefont{Gandhi}},
  \bibinfo{author}{\bibfnamefont{B.}~\bibnamefont{Kayser}},
  \bibinfo{author}{\bibfnamefont{S.}~\bibnamefont{Prakash}}, \bibnamefont{and}
  \bibinfo{author}{\bibfnamefont{S.}~\bibnamefont{Roy}} (\bibinfo{year}{2017}),
  \eprint{1708.01816}.

\bibitem[{\citenamefont{Coloma et~al.}(2017)\citenamefont{Coloma, Forero, and
  Parke}}]{Coloma:2017ptb}
\bibinfo{author}{\bibfnamefont{P.}~\bibnamefont{Coloma}},
  \bibinfo{author}{\bibfnamefont{D.~V.} \bibnamefont{Forero}},
  \bibnamefont{and} \bibinfo{author}{\bibfnamefont{S.~J.} \bibnamefont{Parke}}
  (\bibinfo{year}{2017}), \eprint{1707.05348}.

\end{thebibliography}

\end{document}